\documentclass[%
 reprint,
 amsmath,amssymb,
 aps,
]{revtex4-2}

\usepackage{graphicx}
\usepackage{dcolumn}
\usepackage{bm}
\usepackage{xcolor}
\usepackage{tabularx}

\usepackage{adjustbox}

\newcommand\apjl{ApJ}                
\newcommand\mnras{MNRAS}             
\newcommand\aap{A\&A}                
\newcommand\apjs{ApJS}               
\newcommand\physrep{Reports on Progress in Physics}     

\begin{document}

\preprint{APS/123-QED}

\title{High-Energy Neutrino Fluxes from Hard-TeV BL Lacs}


\author{E. Aguilar-Ruiz$^{1}$}\email{E-mail: eaguilar@astro.unam.mx}
 \author{N. Fraija$^{1}$}%
  \author{A. Galvan-Gamez$^{1}$}
  \affiliation{%
 $^{1}$Instituto de Astronom\'ia, Universidad Nacional Aut\'onoma de M\'exico, Circuito Exterior, C.U., A. Postal 70-264, 04510 M\'exico D.F., M\'exico
}%

\date{\today}

\begin{abstract}

Blazars have been pointed out as promising high-energy (HE) neutrinos sources, although the mechanism is still under debate. The blazars with a hard-TeV spectrum, which leptonic models can hardly explain, can be successfully interpreted in the hadronic scenarios. Recently, Aguilar et al. proposed a lepto-hadronic two-zone model to explain the multi-wavelength observations of the six best-known extreme BL Lacs and showed that the hadronic component could mainly interpret very-high-energy (VHE) emission. In this work, we apply this hadronic model to describe the VHE gamma-ray fluxes of 14 extreme BL Lacs and estimate the respective HE neutrino flux from charge-pion decay products. Finally, we compare our result with the diffuse flux observed by the IceCube telescope, showing that the neutrino fluxes from these objects are negligible. 
\end{abstract}

\maketitle

\section{\label{sec:level1}INTRODUCTION}
%
%
The so-called \textit{extreme} BL Lacs are interesting astrophysical objects to study efficient particle accelerators, which could be linked to the ultra-high-energy cosmic ray (UHECR) and the high-energy (HE) neutrino observations \citep{Biteau2020NatAs...4..124B}. 
There are two ways the extreme behavior in these sources is manifest: as extreme synchrotron peak \citep[EHSPs;][]{2001A&A...371..512C} or/and extreme (hard)-TeV BL Lacs \citep[TBLs;][]{Tavecchio&etal_2011, Costamante2018}. Both conditions are characterized by having an extreme shift of the hump's peak on its spectral energy distribution (SED),  where for the synchrotron hump, the spectral break is above $\gtrsim 2.4 \times 10^{17} \, {\rm Hz}$, and for the HE hump, the energy spectral break is above $\gtrsim 2\, {\rm \, TeV}$ with spectral index characterizing a hard spectrum ($ \gamma_{\rm VHE}^{\rm intr} \lesssim 2 $). An interesting feature is that some EHSP exhibit TBL behavior only during flares activities, suggesting the hard gamma-ray spectrum is not a permanent stage of EHSP \citep[e.g.,][]{Ahnen2018,Ahanorian1999A&A...349...11A, Aleski2014A&A...563A..90A}.

Leptonic models (e.g., electron synchrotron process) successfully describe the low-energy hump of the EHSP and TBL objects without presenting big problems for the standard shock acceleration mechanism with one-zone geometry \citep[e.g., see][]{Ahanorian_2004vhec.book.....A, Tavecchio_2010MNRAS.401.1570T, 2017APh....89...14F, 2017ApJS..232....7F}. However, considering the standard one-zone synchrotron self-Compton (SSC) model, there are a lot of issues trying to model the HE hump of TBLs \citep[see ref.][for a recent review]{Biteau_2020NatAs...4..124B}, mainly because it demands parameters with exceptional and atypical values \citep{Costamante2018, Kaufmann2011,2009MNRAS.399L..59T, 2006MNRAS.368L..52K, Lefa2011}.  Different leptonic models have been proposed as solutions to the issues of the SSC model, for instance, an external radiation field, a Maxwellian-like \cite{Lefa2011}, and an anisotropic electron distribution \cite{Tavecchio&Sobacchi_2020}.  On the other hand, either hadronic or lepto-hadronic models promise to relax the atypical parameter values demanded by the SSC model for TBLs \citep{2015MNRAS.448..910C, Tavecchio_2014, Zheng&etal_2016}. 

The key to discriminating between the models with one or more hadronic components is the neutrino association with TBL, which could be tested with IceCube Observatory at TeV-PeV energies. A total of 102 HE-neutrino events has been reported by IceCube collaboration to the so-called High-Energy Starting Events (HESE) catalog \citep{2013PhRvL.111b1103A,2013Sci...342E...1I,2017arXiv171001191I, IceCube2020arXiv200109520I, IceCube_2021PhRvD.104b2002A}.  Meanwhile, 36 events are reported to the so-called Extremely High Energy (EHE) catalog \citep{IceCube_2016ApJ...833....3A,Icecube_2017ICRC...35.1005H,IceCube_2019ICRC...36.1017S}.

The measured neutrino fluxes using both catalogs show consistent values with single power-law functions, with spectral indexes of $\gamma_{\rm \nu, HESE} = -2.87^{+0.20}_{-0.19}$ and  $\gamma_{\rm \nu, EHE} = -2.28^{+0.08}_{-0.09}$ for HESE and EHE, respectively. On the other hand, when only cascade data for electron and tau neutrinos are considered, the measured flux shows a spectrum with a spectral index of $\gamma_{\rm \nu,bkg} = -2.53 \pm 0.07$ and normalization constant at 100 TeV of $\phi_{\rm \nu,bkg} = 1.66 ^{+0.25}_{-0.27}\, C_0$ with $C_0= 3\times 10^{-18} {\rm \, GeV^{-1} \, s^{-1} \, cm^{-2} \, sr^{-1}}$ \citep{IceCube2020arXiv200109520I}.  Although many astrophysical objects have been proposed as potential sources \citep[for a review see][]{2008PhR...458..173B, 2015RPPh...78l6901A}, the only HE-neutrino source identified up to date is the BL Lac TXS 0506 +056 \cite{2018Sci...361.1378I, 2018Sci...361..147I}.  However, the mechanism of how the HE neutrinos are produced in that blazar is not precise, and different mechanisms have been explored \cite[e.g.,][]{Gao_2019NatAs...3...88G,Xue_2021ApJ...906...51X,Liu_2019PhRvD..99f3008L, Fraija_2020MNRAS.497.5318F,Cerruti_2019ApJ...874L...9H,Halzen_2019ApJ...874L...9H,Kievani_2018ApJ...864...84K,Riemer_2019ApJ...881...46R, Rodrigues_2019ApJ...874L..29R}. 

Some works suggest that BL Lacs are prominent sources to explain the HE-neutrinos reported by IceCube.   For instance,  under the blazar simplified view approach, \cite{Padovani2015MNRAS.452.1877P} concluded that BL Lacs can explain the $\sim 10$ percent of the neutrino background of IceCube above $0.5 \, \rm PeV$.  Further, they suggested that a $\approx 20$ percent level at lower energies could be produced by individual BL Lacs.  Moreover, \cite{Padovani2016MNRAS.457.3582P} found a correlation between the neutrino catalog and EHSP with a $\approx 10-20$ percent of the neutrino signal until that moment.  Their result also showed $\approx$ 6 EHSP as the most likely neutrino candidate.  Interestingly,  \cite{Padovani2014MNRAS.443..474P} previously pointed out TBL H2356-309 as a possible emitter source of the IC10 event.  Following this result, \cite{Petropolou2015MNRAS.448.2412P} proposed a leptohadronic model with feasible parameter values to describe the SED of the neutrino candidates and found that some of these sources could be in agreement with the IceCube observation.  It is worth noting that although recent detection of the HE-neutrino Ice-Cube 200107A was found in temporal coincidence with the flaring activity of EHSP 3HSPJ095507.1+355100 \citep{Giommi2020ATel13394....1G},  there is no evidence of a real connection.

Aguilar-Ruiz et al. \cite{AguilarRuiz_2022MNRAS.512.1557A} has proposed a lepto-hadronic two-zone model to explain the multiwavelength observations in the six best-known extreme BL Lacs. The very-high-energy (VHE) gamma-ray observations are characterized by photo-hadronic activities in a blob near the AGN core and SSC processes in an outer blob. Photo-hadronic interactions occur when accelerated protons in the inner blob collide with annihilation line photons from a sub-relativistic pair plasma.  In this study, we generalize this new scenario to other extreme BL Lacs, by considering the complete catalog reported up to date, which is made up of 14 objects \cite{Costamante2020MNRAS.491.2771C, MAGIC2019MNRAS.490.2284M, MAGIC2020ApJS..247...16A,2021arXiv210802232D},
to describe the TeV gamma-ray spectrum and compute the TeV neutrino counterpart.   To do so, we first model the TeV gamma-ray spectrum of each TBL to find the model's parameter values, and then add the individual neutrino spectrum to get the overall contribution of TBLs.  The following is how this work is organized: Section 2 contains a detailed overview of our model.
In section 3, we present our findings.  Finally, in section 4, we discuss our findings and their implications.   Throughout this work we used the cosmology based on a $\Lambda$CDM model where $H_0 = 67.4 \; \rm km \, s^{-1} \, Mpc^{-1}$, matter energy density $\Omega_M = 0.315$, and dark energy density $\Omega_\Lambda = 1 - \Omega_M $ \citep{2020A&A...641A...6P}.

\section{MODELLING THE EMISSION OF HARD TEV BL LAC}\label{sec:TBL_hadronic_model}

The origin of VHE gamma-rays ($>100 \, \rm GeV$) emission observed in TBL challenges all current models, the formation of a hard spectrum ($\leq 2$) hardly can be accommodated by the one-zone SSC model \cite[see,][for a recent review]{Biteau_2020NatAs...4..124B}. Recently \cite{AguilarRuiz_2022MNRAS.512.1557A} (hereafter AR2022) proposed a two-zone leptohadronic model to explain that VHE emission, they assumed two dissipation regions named as the outer and the inner blob. The outer blob produces the observed flux from X-rays to sub-TeV gamma rays by a SSC model. On the other hand, the inner blob produces observed flux of VHE gamma rays by photopion process. 

Following the treatment of AR2022, we define the reference frames used in this work. They are the observer, the AGN, the pair-plasma, and the blob references frames. The observer and the AGN frames are represented by Latin capital letters, being the only difference between them is the use of the superscript ob
for the observer frame. Meanwhile, the pair-plasma and
blob frames are represented by Greek letters, and we utilize prime to distinguish them in the case of the blob frame. For example, the photon energy in the observer, the AGN, the pair-plasma, and the blob frames will be written as $E^{\rm ob}$ , $E$, $\varepsilon$ and $\varepsilon'$ , respectively.

\subsection{Emitting regions}

We summarize the main features and assumptions of the AR2022 model.
\subsubsection{The outer blob:} 
    The emission of this region is modeled using the one-zone SSC model and considering typical parameters for the blazar. Then we assume that accelerated electrons are dominant over protons, $n_e \gg n_p$; then, only leptonic processes must be considered.   
    We assume this region moves with relativistic speed, with Lorentz and Doppler boost factors set as $\Gamma_o = 5$ and $\mathcal{D}_o \simeq 10$, respectively. 
    Finally, we assume this blob is localized at the end of the acceleration and collimation zone, i.e., $r_o\sim 10^3 \,R_g$ \citep[e.g,][]{Marscher_2008,Walker_2018ApJ...855..128W}, then the blob's size is constrained using the relation $r_o \simeq 2\Gamma_o R_o'' \sim 10^{17} \, \rm cm$ with $R_g= GM_\bullet/c^2 \sim 10^{14} \, \rm cm$ the Schwarzschild radius for a SMBH mass of $M_{\bullet} = 10^{9} M_{\odot}$. Therefore, we obtain $R_o'' \sim 10^{16} \, \rm cm$.
    
%
%
%
\subsubsection{Inner Blob}
The motivation to consider an inner blob where TeV gamma-rays are produced arises from the fact that an efficient photopion process requires the following condition
\begin{equation*}
    L_X'\sim 10^{45} \, {\rm erg \, s^{-1}} \, \left(\frac{f_\pi}{10^{-1}}\right) \, 
    \left(\frac{R_b'}{10^{14} \, \rm cm}\right) \left(\frac{\varepsilon_X'}{100 \, \rm keV}\right) \,.
\end{equation*}
Therefore, to reduce the luminosity requirement, compact regions and low values of boost factor are favored; the condition may be fulfilled in the inner jet zones, inside the acceleration and collimation zone, $r_i \lesssim 10^3 \, R_g$. Note that if a region emits TeV gamma-rays as the outer blob, the required luminosity is significantly higher than the above result, i.e., $L_X^{\rm ob}= \mathcal{D}^4 L_X' \sim 10^{51} \rm \, erg \, s^{-1}$, which is an unfeasible value for extreme BL Lacs \cite{Costamante2018,Costamante2020MNRAS.491.2771C}. Furthermore, to produce VHE gamma-rays by photopion process is necessary the existence of seed photons with energies of
\begin{equation*}
    \varepsilon_X' \gtrsim 100 \, {\rm keV} 
    \, \mathcal{D} 
    \, \left(\frac{E_\gamma^{\rm ob}}{0.1 \, \rm TeV}\right)^{-1} 
    \, .
\end{equation*}
Nevertheless, if the spectrum of seed photons extends to lower energies below $\sim 1 \, \rm keV$, TeV gamma rays hardly may escape from the source. On the other hand, if we consider the spectrum's shape like an emission line, the VHE emission should escape if the line emission is centered at energies greater than $\sim 100 \, \rm keV$. In this scenario, only MeV-GeV gamma rays must be attenuated. Therefore, the model of AR2022 considers the annihilation line emission centered at $\sim 511 \, \rm keV$. This line emission has been suggested to arise in compact regions near the SMBH \citep[e.g., ][]{1999MNRAS.305..181B}.

The inner blob is assumed to have one electron per proton, $n_e=n_p$, and the same spectral index for electrons and protons $\alpha_e = \alpha_p$. If the magnetic field luminosity is conserved along the jet, the magnetic field in two blobs are related as $B_i = (r_o/r_i) ( \Gamma_o/\Gamma_i)\, B_o$ \cite[e.g.,][]{Osullivan_2009MNRAS.400...26O, Konigl_1981ApJ...243..700K}, then the magnetic field must be stronger than in the outer blob. Stronger values of $B_i$ must suppress electron acceleration, then synchrotron cooling avoiding electrons reaches significantly higher energies, $\gamma_e \sim 7$ for $B_i\sim 100 \, \rm G$. Their flux produced by primary electrons is small and only has a signature at radio frequencies. 
On the other hand, protons are not efficiently cooling down by synchrotron and can reach higher energies. We assume protons are accelerated by Fermi shock acceleration, and then a harder spectrum can be considered even with a spectral index of $1.5$ \cite{Stecker2007ApJ...667L..29S}.

\subsection{TeV gamma-rays and HE-neutrinos production}
We briefly review the basic properties of the model to produce TeV gamma-rays considering photohadronic processes. 
We focus on the inner blob (hereafter named only as blob) and the pair plasma, which explain the VHE gamma-rays and produce the neutrino emission.

In our model the blob moves with speed, $\beta_{\rm b}$, and Lorentz factor, $\Gamma_{\rm b}$, respectively. The viewing angle is close to the line of sight, $\theta_{\rm obs}\lesssim 1/\Gamma_{\rm b}$, then the boost Doppler factor is approximated by $\mathcal{D} \simeq 2\Gamma_{\rm b}$. On the other hand, the pair-plasma moves with sub-relativistic speed, $\beta_{\rm pl}$ and Lorentz factor, $\Gamma_{\rm pl}$. Therefore, the relative Lorentz factor between the blob and the pair plasma is given by
\begin{equation}\label{eq_GammaRel}
\Gamma_{\rm rel} = \Gamma_{\rm b} \Gamma_{\rm pl} \left( 1 - \beta_{\rm b} \beta_{\rm pl} \right)\,.
\end{equation}

The blob's total energy is the sum of the contribution of particles (electrons and protons), photons, and magnetic fields. But for practice, here, we only consider the proton energy contributions. Then the proton bolometric luminosity of the blob in the AGN frame is \cite[e.g.;][]{Celotti&Ghisellini_2008, Bottcher&etal_2013}
\begin{equation}\label{eq_proton_luminosity}
    L_{\rm p} = \pi {R'}_{\rm b}^2 \, \Gamma_{\rm b}^4 \beta_{\rm b} c \int \varepsilon'_{\rm p} N'_{\rm p}(\varepsilon'_{p}) d\varepsilon'_{\rm p} \, , 
\end{equation}
where $R'_b$ is the size of the blob and $N_p'$ is the isotropic and homogeneous proton distribution, which follows a power-law given by 
\begin{equation}\label{eq_pDistribution}
{N_p'}({\varepsilon_p'}) = {K_p'} \left( \frac{\varepsilon'_p}{m_p c^2} \right)^{-\alpha_{p}} \, 
\qquad \varepsilon'_{p, \rm min} \leq \varepsilon_p' \leq \varepsilon'_{p, \rm max}\,,
\end{equation}
where $K_p'$ is the normalization constant, $\alpha_p$ is the proton spectral index and ${\varepsilon}'_{p, \rm min}$ and ${\varepsilon}'_{p, \rm max}$ correspond to the minimum and maximum energy in the comoving frame, respectively.

%
%
\subsubsection{Annihilation line}
We consider the existence of a pair plasma that produces an annihilation line with a peak at $\varepsilon_{\rm pl} = 511 \, \rm keV$ \citep{1999MNRAS.305..181B}. The photon distribution's shape is assumed very narrow such that it can be described in delta approximation as \citep{Fraija_2020MNRAS.497.5318F}

\begin{equation}\label{eq_line_distribution}
    n_{\rm pl} (\varepsilon) = \frac{u_{\rm pl} }{\varepsilon_{\rm pl}}\, \delta(\varepsilon - \varepsilon_{\rm pl})\,,
\end{equation}

where $u_{\rm pl} = L_{\rm keV} / (\Omega_{\rm pl} R_{\rm pl}^2 \beta_{\rm ph } c)$ is the energy density of photons produced by the pair-plasma. Photons escape from the pair-plasma at the photosphere region, which moves  with velocity $\beta_{\rm pl} \approx 0.3-0.5$ \citep{1999MNRAS.305..181B}, and radius of the $R_{\rm ph} \sim R_g$ where $R_g=G M_{\bullet}/c^2$ is the Schwarzschild radius with $G$ the gravitational constant and $M_{\bullet}$ is the mass of the supermassive black-hole (SMBH). The emission of the annihilation line is very anisotropic and the the main contribution lies inside $\Omega_{\rm pl} \lesssim 0.2\pi$. Furthermore, in order to guarantee the formation of the pair-plasma the luminosity above $\rm 511 \, keV$ must be greater than $L_{\rm keV} \gtrsim 3 \times 10^{-3} \, L_{\rm E}$, where $L_E \approx 1.26 \times 10^{47} \, {\rm erg\,s^{-1}}$ is the Eddington luminosity for a SMBH of $M_{\bullet}=10^9 M_{1\odot}$. This condition arises as a consequence of producing an optical depth environment to have efficient pair production, which occurs at compactness of $\l > 30$ \cite{1999MNRAS.305..181B}.
Here we assume that $L_{\rm keV}$ is the order of the disc luminosity, $L_d \sim L_{\rm keV}$ which the condition lies in the range of possible values for BL Lacs, $L_d \lesssim 5 \times 10^{-3} L_E$ if the BLR luminosity is $L_{\rm BLR} = 0.1 L_d$ \cite{Ghisellini&etal_2011}. However, if the reprocessed fraction by the BLR is smaller than 10 percent of the luminosity disk, it can even be higher.

\subsubsection{Photopion process}
The co-existing  of the annihilation line radiation field close to the inner blob produce interactions with accelerated protons inside it via the photopion process \citep{Fraija_2020MNRAS.497.5318F,AguilarRuiz_2022MNRAS.512.1557A}. 
The emission of the pair plasma is seen with redshift effect in the blob frame, then the energy and energy density given by \citep{Dermer&Menon2009}
\begin{equation}\label{eq_plasma_energy}
   \varepsilon^{\prime}_{\rm pl} \simeq {\varepsilon_{\rm pl}}/(2\Gamma_{\rm rel}) \, , \qquad   u^{\prime}_{\rm pl} = \frac{u_{\rm pl}}{\Gamma_{\rm rel}^2 (1 + \beta^2/3)}\,.
\end{equation}
%
%
A simple estimation of the energy of gamma-rays produced by the photopion process with photons of the annihilation line could be estimated using the delta-resonance channel $p+\gamma \rightarrow \Delta^+$, then from relativistic kinematics 
\begin{equation}\label{eq_photopion_gamma_rays}
E_\gamma \gtrsim 75 \, {\rm GeV} \, \Gamma_{\rm rel} \, \Gamma_{\rm b} \, ,
\end{equation}
%
%
this coincides with the energy range of gamma-rays observed in TBLs.
The spectrum of gamma-rays and neutrinos produced by the decay of neutral and charged pions, respectively, are calculated using the treatment of \citep{Kelner&Aharonian_2008}
\begin{align}\label{eq_photopion_flux}
{Q'}^{p\pi}_{\gamma,\nu}(\varepsilon'_{\gamma,\nu}) 
&= \int_{\varepsilon'_{\rm min}}^{\varepsilon'_{\rm max}}                                 \frac{d\varepsilon'_p}{\varepsilon'_p} N'_p(\varepsilon'_p) \,
   \int_{\varepsilon'_{min}}^{\infty}
      d\varepsilon' \, n'_{\rm pl}(\varepsilon') \, \Phi_{\gamma,\nu} (\eta,x),
\end{align}
where the function $\Phi_{\gamma,\nu}$ is parametrized by the authors and they define the parameters $\eta=\frac{4\varepsilon'_p\varepsilon'}{m_p^2 c^4}$ and $x=\frac{\varepsilon'_{\gamma, \nu}}{\varepsilon'_p}$. 

The photopion timescale is calculated using the expression \citep{Dermer&Menon2009}
\begin{align}\label{eq_photopion_loss_timescale}
{t'}_{p\pi}^{-1} = \frac{c}{2\gamma_p^2} \int_{\frac{\varepsilon_{\rm th}}{2\gamma_p}}^\infty d\varepsilon' \frac{n_{\rm pl}(\varepsilon')}{\varepsilon'^2} \int_{\varepsilon_{\rm th}}^{2\gamma_p\varepsilon'} d\varepsilon_r \varepsilon_r \sigma_{p\pi} (\varepsilon_r) K_{p\pi}(\varepsilon_r) \, ,
\end{align}
where $\varepsilon_{\rm th} \approx 150 \, \rm MeV$ is the threshold energy, $\varepsilon_r$ is the photon energy in the proton rest-frame, $\sigma_{p\pi}$ is the photopion cross-section and $ K_{p\pi}$ is the inelasticity coefficient.  Here, we use the two-step approximation, $\sigma_{p\pi} K_{p\pi} = 68 \, {\rm \mu barn}$ for $199 \, {\rm MeV} \leq \varepsilon_r \leq 500 \, {\rm MeV}$ and $\sigma_{p\pi} K_{p\pi} = 72 \, {\rm \mu barn}$ for $\varepsilon_r > 500 \, {\rm MeV}$.

Similarly, the energy loss timescale due to fotopair process, $p+\gamma \rightarrow p + e^\pm$, is calculated as 
\begin{align}\label{eq_pe_loss}
{t'}_{pe}^{-1}(\varepsilon'_p) = 
\frac{3 \sigma_{\rm T} \alpha_f m_e^3 }{16 \pi} \frac{m_p c^2}{ {\varepsilon'_p}^2} 
\int_{\varepsilon'_{\rm th}}^\infty d\varepsilon' \,  \frac{n'_{\rm pl}(\varepsilon')}{{\varepsilon'}^2} \varphi \left( \frac{2 \varepsilon'_p \varepsilon'}{m_p m_e c^4} \right)\, ,
\end{align}
where $\varepsilon'_{\rm th} \simeq 2m_e m_p c^4/\varepsilon'_p$ is the threshold energy, $\alpha_f$ the fine structure constant, and the function $\varphi$ is parametrized following  \cite{Chodorowski_1992}.

All quantities described by equations (\ref{eq_photopion_flux}-\ref{eq_pe_loss}) are valid for isotropic photon distributions. However, the radiation field produced by the pair plasma is very anisotropic \cite{1999MNRAS.305..181B}. 
Moreover, further beamed effects are produced by relativistic conditions doing photons measured in the inner blob being more anisotropic. If photons are emitted very close to the movement direction of the pair plasma and the inner blob, the emission cannot be integrated over the total polar angle cover, $\mu=-1$ to $\mu=1$. Therefore, the flux must be a factor of $< 2$ than the isotropic case. 
Additionally, the number density is boosted in the inner blob, as $n_{\rm pl}' = n_{\rm pl}/(2\Gamma_{\rm rel})$ for an isotropic and mono-energetic distribution case, as the equation 1,  and for a mono-energetic distribution with photons emitted in a particular direction, $\mu=1$, the number density transformation is $n_{\rm pl}' = n_{\rm pl}/(\Gamma_{\rm rel}(1+\beta_{\rm rel}))$.  Therefore, this correction factor is not relevant.

\subsection{Neutrino diffuse flux}
The neutrino diffuse flux produced by the TBL population can be calculated using the gamma-ray luminosity function (GLF) of BL Lacs and integrating over redshift, Luminosity and spectral index \citep[e.g.,][]{ Fraija_2020MNRAS.497.5318F, Becker_2014PhRvD..89l3005B, Becker_2005APh....23..355B}
\begin{multline}\label{eq_diffuse_neutrino}
\phi_\nu (E_\nu) =
\int_0^{z_{\rm max}} dz
\int_{L_{\gamma, \rm min}}^{L_{\gamma, \rm max}} dL_\gamma  \, 
\int_{\gamma_{\rm int, \rm min}}^{\gamma_{\rm int, max}}
d\gamma_{\rm int} 
\\
\rho_{\gamma}(L_\gamma, z) \times
\frac{dV_{\rm c}}{dz} \times \frac{d N}{d\gamma_{\rm int}} 
\times\frac{Q_{\nu} (E_\nu)}{4\pi d_L^2} \, ,
\end{multline}
where $\rho_\gamma$ is GLF which represent the numbers of BL Lacs per comoving volume in the range of luminosities  [$L_\gamma + L_\gamma dL_\gamma$] \citep{Qu_2019MNRAS.490..758Q, Zeng_2014MNRAS.441.1760Z},  $dV_{\rm c}/dz = cd_L^2 / (H_0 (1+z)^2 \sqrt{(1+z)^3 \Omega_m + \Omega_\Lambda})$ is comoving volume element as function of the redshift per angle solid unit, and $dN/d\gamma_{\rm int}$ is the distribution of intrinsic spectral indexes.

\section{RESULTS}\label{sec:TBL_Application}

Up to date, 14 BL Lacs have been detected having an intrinsic hard-spectrum in the TeV gamma-ray band, i.e., $\gamma_{\rm VHE}^{\rm int} \leq 2$.  They are listed in Table \ref{tab_EHSP_list}. If the uncertainties are considered, only six of them can be tagged as TBL, i.e, 1ES 0229 +200, RGB J0710 +591, 1ES 0347-121, 1ES 1101-232, 1ES 1218 +304, TXS 0210 +515. But if instead, we take into account the best-value without uncertainty,  we can add three more sources to this list, i.e., 1ES 0414 +009, H 2356-309, and 1ES 1426 +428. The rest could be considered cases that lie at the limit of the transition between soft-to-hard TeV BL Lacs ($\gamma_{\rm VHE}^{\rm int} \approx 2)$; i.e.,  PKS 0548-322, 1ES 2037 +521, RGB J2042 +244, 2WHSP J073326.7+515354 and 1RXS J195815.6-301119. From this five latter, it is important to mention the case of RGB J2042 +244 for which only a hint signal was reported and could not be tagged as a real detection \citep{MAGIC2020ApJS..247...16A}. In this work we consider all TBL mentioned and enlisted in table \ref{tab_EHSP_list} including the case of RGB J2042 +244.   We take the observed spectrum of VHE gamma-ray energies of each TBL after the corrections due to the extragalactic background light (EBL) absorption effect. This research posits that VHE gamma-rays have a hadronic origin and are created by a photopion process, as outlined in the preceding section. As a result, a neutrino flux must be created in addition to the VHE gamma-ray flux.
To determine the neutrino flux associated with each TBL, we first model the VHE gamma-ray spectrum including the entire broad band emission SED using the prescription shown in section \ref{sec:TBL_hadronic_model}.

\subsection{Modelling broadband emission}
We model the broadband SED of eight TBLs in addition to the previous six ones treated by AR2022, the complete catalog are listed in Table \ref{tab_EHSP_list}. Respect to the outer blob, we follow the same treatment as AR2022 and you cad see there in for more details. Moreover, for simplicity and because is not the main goal of this work, we do not consider the external Compton scattering of BLR/DT photons in the outer blob.  Then we expect that $r_{o}>r_{\rm BLR, DT}$.
Therefore we only focus on the inner blob treatment where the neutrino emission is produced. Therefore, hereafter, for simplicity we refer to the inner blob only as the blob.


To apply our photo-hadronic model, we begin setting the same parameters values of the pair-plasma for all TBLs.  To guarantee the formation of a pair-plasma  we establish the luminosity above 511 keV as $L_{\rm keV} \approx 5 \times 10^{-3} L_E$. This option is justified by assuming that this luminosity is equal to the luminosity of the disc. Furthermore, in the broad-line zone, the value of choice is the highest allowed by a BL Lac.  This luminosity represents ten percent of the luminosity of the disc, i.e., $L_d = 10 L_{\rm BLR}$ \cite{Ghisellini&etal_2011} \footnote{It is the value suggested as a transition one between the BL Lac objects to flat-spectrum radio quasars}. The photosphere velocity of the pair-plasma is  $\beta_{\rm ph} \approx 0.3$, which corresponds to a Lorentz factor of $\Gamma_{\rm pl} \approx 1.05$. Therefore, the pair-plasma produces annihilation photons with energy density of $u_{\rm pl} \approx 3 \times 10^{6} \, \rm erg \, cm^{-3}$  for $M_{\bullet} = 10^9 M_{\odot}$.

The blob parameters determine the resulting spectrum at the VHE gamma-ray band, dependent on four parameters: the boost Lorentz factor, the size of the blob, the proton spectral index, and the proton normalization constant. Here, we restrict the values using simple estimations as follows described. 
Using equations \ref{eq_photopion_gamma_rays} and \ref{eq_GammaRel}, we can note that the peak of the spectrum depends on the value of the $\Gamma_{\rm pl}$ and $\Gamma_{\rm b}$. For instance, if we consider a blob's Lorentz factor of $\Gamma_b = 1.5 \, ( \beta_{b}\approx 0.75)$ the gamma-ray energy peak becomes
\begin{equation}\label{eq_gamma_peak}
    E_\gamma \gtrsim 0.177 \, {\rm TeV} \, ( \Gamma_{\rm pl} / 1.05 ) \, ( \Gamma_{\rm b}/1.5 )^2 \, .
\end{equation}
Note that the gamma-ray peak shows more dependency with the value of $\Gamma_{\rm b}$ instead of $\Gamma_{\rm pl}$, even if this latter were a free parameter, it only should take values in a small range $\beta_{\rm pl} = 0.3-0.7 (\Gamma_{\rm pl} = 1.05-1.4)$ and then not play an important role. In our study, all TBLs have a peak above $>2 \, \rm TeV$,  therefore only values of $\Gamma_{\rm b} \gtrsim 1.5$ should be expected. For example, if $\Gamma_{\rm b}$ has a value as high as 10, the gamma-ray peak is shifted at $\varepsilon_\gamma \sim$ 40 TeV, which leave a difficulty to fit the observations. Note that the TBL with greatest gamma-ray peak  $\varepsilon_\gamma \approx 10 \, \rm TeV$ is 1ES 0229 +200, and even this object hardly demands greater values than $\Gamma_{\rm b} \approx 5$. Taking this argument, we expect that only mildly-to-relativistic blob's Lorentz factors must be favoured. Concerning the size of the blob, we adopt values $R'_b \sim {\rm few} \times\, R_g$. 

\begin{figure}
\centering
\includegraphics[width=\linewidth]{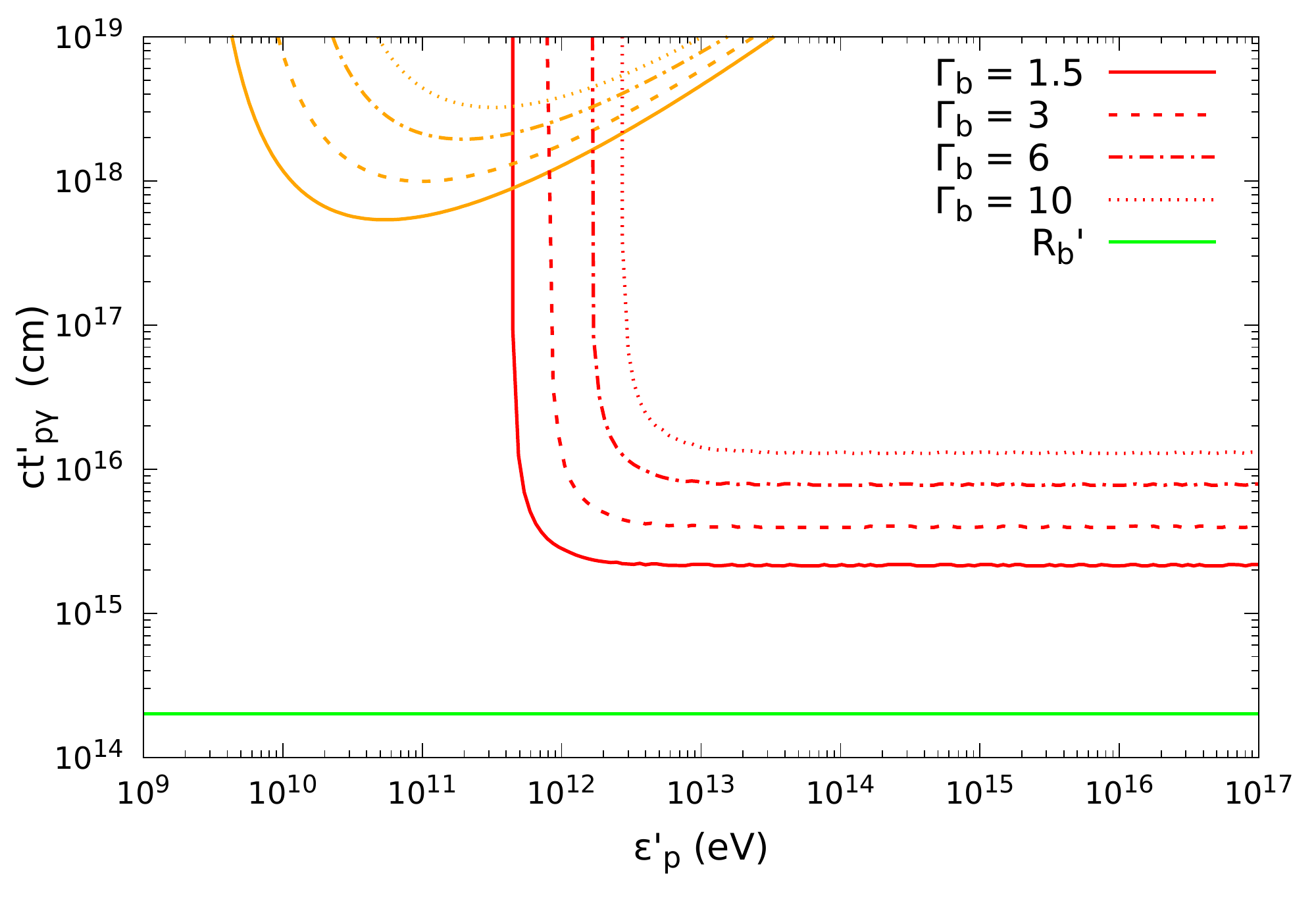} 
\caption{Proton timescales in the comoving frame for photopion (red lines) and photopair (orange lines) processes. We consider the seed photons those coming from the pair-plasma. Furthermore, we show the typical size of the inner blob, $R_b'=2\times10^{14} \, \rm cm$ (green line). We assume a SMBH mass of $M_\bullet = 10^9 M_\odot$, an annihilation line luminosity $L_{\rm keV} =  5\times10^{-3}L_{\rm Edd}$, a pair-plasma velocity of $\beta_{\rm pl}=0.3$ and the boost Lorentz factor of the blob $\Gamma_{i}=1.5, 3, 6, 10$.}
\label{fig_timescale}
\end{figure}

Using equation \ref{eq_photopion_loss_timescale} and \ref{eq_pe_loss}, we calculate the proton loss timescale due to the photopion and photopair processes for different values of $\Gamma_{b}$; the result is plotted in Figure \ref{fig_timescale}. This figure shows that the process only occurs in a given range of proton energies which is determined by the value of $\Gamma_{b}$.  While we increase the value of $\Gamma_{b}$, the value of the timescale also increases, and the energy threshold is shifted to higher energies conserving the energy range's width. As we previously mentioned, we do not expect values greater than $\Gamma_{b}\approx 5$.  Furthermore, note that the photopion timescale has a quasi-flat shape (see Figure \ref{fig_timescale}), then the photopion efficiency, $f_{p\pi} \simeq (R'_b/c) t_{p\pi}^{-1}$ must be almost constant. Note that, for a blob's size of $R'_b \sim 3 \times 10^{14} \, \rm cm$, the maximum photopion efficiency reached is $f_{p\pi}\sim 0.1$ while the photopair efficiency is only $f_{pe} \sim 10^{-3}$.
The spectral shape of gamma rays could be well traced by the shape of the proton distribution. Due to this, here we take the intrinsic spectral index of VHE gamma-rays as the values of the proton spectral index, $\gamma^{\rm intr}_{\rm VHE} =\alpha_p$.


Due to the value of $\varepsilon'_{p, \rm max}$ determines the peak of the gamma-ray spectrum, in our model, protons must be accelerated beyond $\varepsilon'_p \gtrsim 120 \, (E_\gamma^{\rm ob}/12 \, {\rm TeV}) \, \mathcal{D}^{-1}$ to obtain $E_\gamma \sim 12 \, \rm TeV$; at least for the case of 1ES 0229+200 (see table \ref{tab_EHSP_list}). Therefore, assuming the value of $\varepsilon_{p, \rm max}' \gtrsim 100 \, \rm TeV$ must be enough to explain the VHE gamma-ray data.
Nonetheless, by comparing $t'_{\rm acc} = \varepsilon_{p}'/ (\eta eB'c)$ with the adiabatic loss $t'_{\rm ad}= R'/c$, we obtain that the maximum proton energy reached by Fermi mechanism is
\begin{equation}\label{eq_Epmax}
    \varepsilon_{p, \rm max}' \sim 300 {\, \rm PeV} \left( \frac{\eta}{0.1} \right) \, \left( \frac{B'}{100 \, \rm G}\right) \left( \frac{R'}{2 \times 10^{14} \, \rm cm}\right)\,,
\end{equation}
where 
$\eta$ is the acceleration efficiency. These values indicate that protons are accelerated beyond the energies to photopion works in our model. 
In this work in order to minimize the power requirement can set the maximum proton energy as $\varepsilon'_{p, \rm max} = 100 \, \rm TeV$,
we do not take higher values, as indicate the equation \ref{eq_Epmax}, because the peak of VHE gamma-rays spectrum is not well constrain and our results only must changes for the neutrino spectrum and for the power requirements. However, for the calculation of diffuse neutrino flux, we consider the impact of $\varepsilon_{p, \rm max}'$ on the resulting spectrum and on the peak of the diffuse spectrum (see section \ref{sec:neutrino_results}).

It is important to mention that in this case proton synchrotron is not a relevant process, because of the losses timescales, $t_{p,syn} \approx 4.5 \times 10^6 \, {\rm sec} \, (B'/100 \, {\rm G})^{-2}
\, (\varepsilon_p'/100 \, {\rm PeV})^{-1}$ is much greater than the dynamical timescale $R_b'/c \approx 6.7 \times 10^{3} \, \rm sec$. Similarly, before decay into other particles, secondary unstable charged particles produced by photopion processes, might cool down by synchrotron process; a simple estimation can be done by comparing the synchrotron loss timescale with the particle lifetime, $t_{lt, \pi^{\pm} (\mu)} \sim E_{\pi^{\pm} (\mu)}/(m_{\pi^{\pm} (\mu)} c^2) \tau_{\pi^{\pm} (\mu)}$ where the mean lifetime for muons is $\tau_{\mu} = 2.197 \times 10^{-6} \rm \, s$ and for charged pions is $\tau_{\pi^{\pm}} = 2.6 \times 10^{-8} \rm \, s$ \cite{ParticleDataGroup_2020PTEP.2020h3C01P}, then we obtain that these particles are not efficiently cool down by synchrotron for energies below $E_{\pi^\pm} \lesssim 1 \, \rm EeV$ and $E_{\mu^\pm} \lesssim 59 \, \rm PeV$ for pions and muons, respectively. Therefore, even if we take higher values than the estimated by equation \ref{eq_Epmax} the synchrotron emission of protons, pions and muons can be neglected.
\textbf{}


Concerning secondary pairs, Figure \ref{fig_timescale} shows that the luminosity of secondary electrons produced by Bethe-Heitler (pe)  is lower than those produced by the photopion process by a factor of $8f_{pe}/f_{p\pi} \sim 10^{-2}$.  An estimation of the energy peak of secondary pairs could be done using the proton energy at the peak's efficiency.  From Figure \ref{fig_timescale}, we can note that the process efficiencies reach the maximum at proton energy  around ${\varepsilon'}_{p,pe}^{\rm pk} \sim 100 \, \rm GeV$ and ${\varepsilon'}_{p,p\pi}^{\rm pk} = \varepsilon'_{p,\rm max} = 100 \, \rm TeV$. And if protons transfer a fraction energy to electron as ${\varepsilon'}_{e,pe} \approx (m_e/m_p) {\varepsilon'}_{p} $ and ${\varepsilon'}_{e,p\pi} \approx 0.05 {\varepsilon'}_{p}$, then electrons are created with Lorentz factor around $\gamma'_{e,pe} \sim 10^2 $ and $\gamma'_{e,p\pi} \sim 4 \times 10^5$, respectively. This correspond to synchrotron emission of
secondary pairs peak at $\varepsilon'_{pe, \rm syn} \sim 0.01 \, {\rm eV} \, (B'/100 \, {\rm G}) (\gamma'_{e,pe}/10^2)^2$ and $\varepsilon'_{p\pi, \rm syn} \sim 1 \, {\rm MeV} \, (B'/100 \, {\rm G}) (\gamma'_{e,p\pi}/ 1\times 10^6)^2$, respectively. With the strong magnetic field such as in the inner blob; $B \sim 100 \, \rm G$, the synchrotron emission of secondary pairs can be estimated as $L_{\rm syn}^{p\pi (pe)} \sim C(ct_{p\pi(pe)})^{-1} R' L_p$, where $C$ corresponds to 1/8 and 1 for photopion and photopair process, respectively (see \citep{Fraija_2020MNRAS.497.5318F, Petropoulu_2015MNRAS.447...36P}). However, as we mention above this contribution could be neglected ($L_{\rm syn}^{pe} \lesssim 10^{-3} L_{\rm syn}^{p\pi}$).  Figure \ref{fig_timescale} shows that for low values of $\Gamma_b=1.5$, the photopion efficiency becomes $f_{p\pi}\sim 0.1$. 
The VHE emission of powerful TBLs is $L_{\gamma}^{\rm VHE} \sim 10^{45} \, \rm erg \, s^{-1}$, which would imply a proton luminosity of $L_p \sim 10^{46} \, \rm erg \, s^{-1}$ and synchrotron luminosity of secondary pairs of $L_{\rm syn}^{p\pi}
\sim 2 \times 10^{44} \, \rm erg \, s^{-1}$.
This emission collides and annihilates with gamma-rays above $E_\gamma \approx 2 \mathcal{D}_i m_e^2 c^4/\varepsilon'_{p\pi, \rm syn} \sim 0.5 \, {\rm MeV} \mathcal{D}_i (\varepsilon'_{p\pi, \rm syn}/1 \, {\rm MeV})^{-1}$. This value is close to the produced by the photons of the pair-plasma (see Figure 2b in \citep{AguilarRuiz_2022MNRAS.512.1557A}). The optical depth is approximately $\tau_{\gamma\gamma} \approx 0.2\sigma_{\rm T} L_{\rm syn}^{p\pi} {R'}_b^{-1}/(4\pi \mathcal{D}_i^4 c)$, then we obtain
\begin{multline}
    \tau_{\gamma\gamma} \approx 0.01 \left( \frac{\mathcal{D}_i}{3} \right)^{-4} \left( \frac{L_{\rm syn}^{p\pi}}{10^{44} \, \rm erg \, s^{-1}} \right) \left( \frac{\varepsilon'_{p\pi, \rm syn}}{1 \, {\rm MeV}} \right)^{-1} 
    \\
    \left( \frac{R'_b}{2\times 10^{14} \, \rm cm} \right)^{-1} \, ,
\end{multline}
and therefore, secondary pairs emission does not attenuate gamma-rays.

Finally, gamma-rays generated inside the blob escape and travel throughout the broad-line region, dusty torus, and maybe an outer blob before emerging. Here, the attenuation effects of these regions due to $\gamma\gamma$ absorption are not considered because this should have a weak or neglected effect at TeV energies (see figure 2b in \cite{AguilarRuiz_2022MNRAS.512.1557A}).

Finally, we perform the fit requiring the MINUIT algorithm \cite{James&Ross1975CoPhC..10..343J, Fraija_etal2021ApJ...918...12F} through the {\it iminuit} \footnote{https://github.com/scikit-hep/iminuit} Python interface. We require the interpolation of the solution with the {\it interp1d} function from the {\it scipy.interpolate} Python object, and the $\chi^2$ regression function using the {\it Chi2Regression} object from the {\it probfit} Python interface. We fix a set of initial parameter values to minimize the function with the {\it iminuit} interface and the {\it migrad} optimizer.
The best-fit, the assumed and the derived values used to model each TBL are listed in Table \ref{tab_model_parameters}, and the respective SEDs are plotted in Figure \ref{fig_VHEspectrum}. For a completely different set of parameters, the same conclusions can be obtained, as the equations are degenerate in these values. As a result, our solution is not the only one possible.

\subsection{Contribution of TBLs to diffuse neutrino flux}\label{sec:neutrino_results}
\begin{figure}
\begin{minipage}[b]{\linewidth}
\centering
\includegraphics[width=\linewidth]{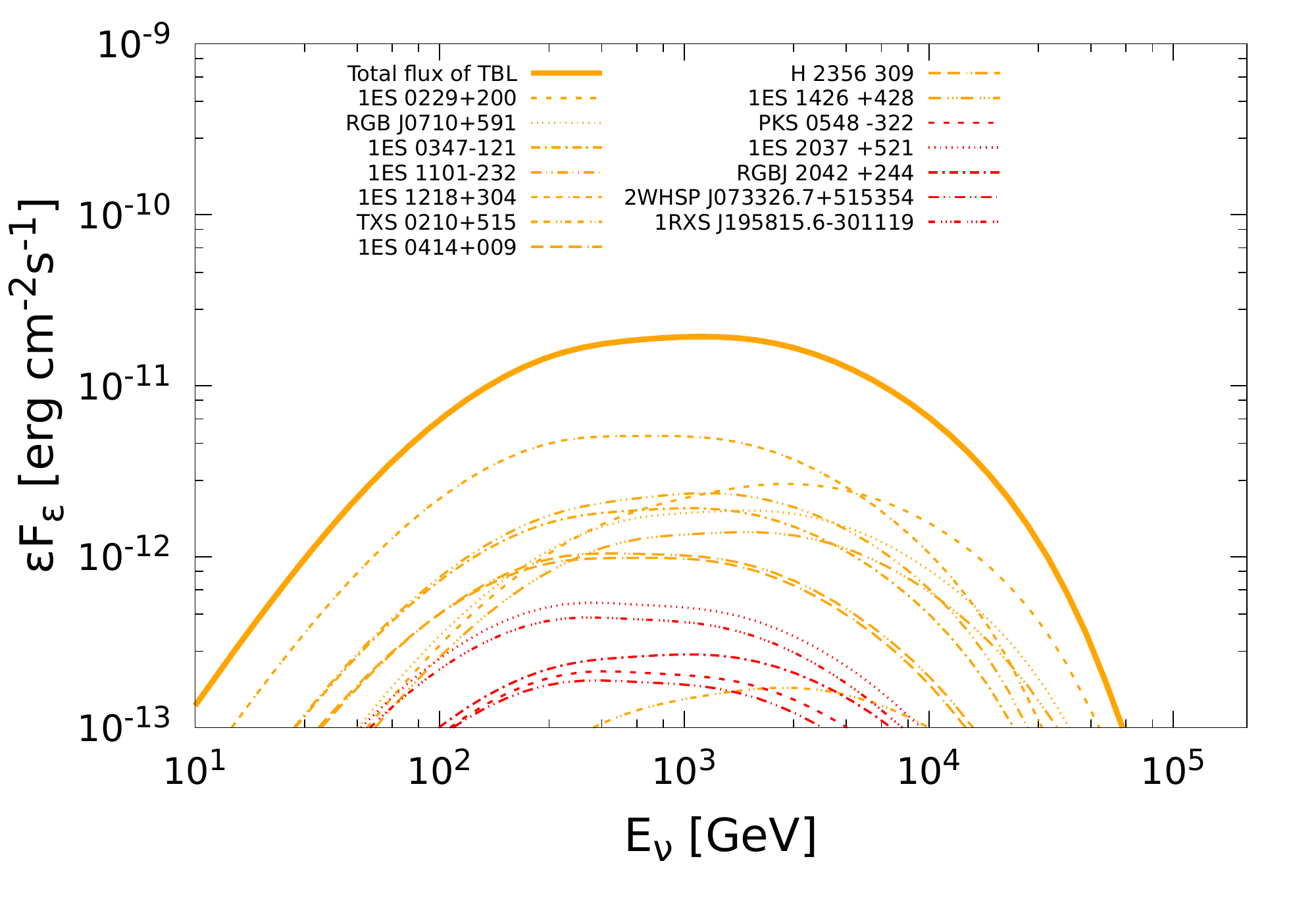} 
\end{minipage}
\caption{The neutrino flux estimated with our photo-hadronic model for each hard-TeV BL Lac. The contribution from TBL objects with a  spectral index value $\gamma^{\rm int}_{\rm VHE} < 2$ are shown in blue line and those ones, which could lie, $\gamma_{\rm VHE}^{\rm intr} \approx 2$,  in the transition regime are shown in red lines.}
\label{fig_Fneutrinos}  
\end{figure}

Once the VHE gamma-ray spectrum is modeled, we use the same parameters to calculate the neutrino spectrum for each TBL. 
Then, the total flux produced by the TBL population is calculated by summing the flux obtained from each object and the result is plotted in Figure \ref{fig_Fneutrinos}.
Finally, we use as benchmark values the parameters found with our model for each TBL, to calculate the diffuse flux produced by TBL using the GLF of BL Lacs (see equation \ref{eq_diffuse_neutrino}). 
%
%
%
In our scenario, the neutrino flux is obtained by integrating over redshift, gamma-ray luminosity, and intrinsic spectral index of gamma rays. We integrate the redshift from $z_{\rm min}=0$ to $z_{\rm max}=2$. This maximum value is an excellent choice to include all contribution of BL Lacs because, beyond $z\sim 1$, the numbers of BL Lacs drops drastically down \cite[see,][]{Zeng_2014MNRAS.441.1760Z}.
In lack of a gamma-ray luminosity function for TBL, we use the case of Pure Evolution Density (PED) function for all BL Lacs given by \citep{Qu_2019MNRAS.490..758Q, Zeng_2014MNRAS.441.1760Z}, and we integrate it from $L_{\gamma,\rm min} = 10^{40} \, \rm erg \, s^{-1}$ to $L_{\gamma,\rm max} = A_\gamma L_{\gamma,\rm max}^{\rm VHE}$, where $L_{\gamma,\rm max}^{\rm VHE}$ is the maximum luminosity in the VHE-band. Table \ref{tab_EHSP_list} shows that the VHE peak for most luminous TBLs is $\gtrsim 10^{45} \rm erg \, s^{-1}$. 
The parameter $A_\gamma$ relates the luminosity of the Fermi energy band (100 MeV - 100 GeV) with the luminosity in the VHE band ($>$100 GeV); then, by assuming that all TBL have the same or similar spectral index in both bands \cite[see,][]{Costamante2018}, we obtain the relation
\begin{equation}
    A_\gamma = \frac{L_{\gamma}}{L_{\gamma, \rm VHE}} = \frac{10^{3(2-\gamma_{\rm int})}-1}{(E_{\gamma,\rm max}^{\rm VHE}/100 \,\rm MeV)^{2-\gamma_{\rm int}}} \, , \quad \gamma_{\rm int} \neq 2 \, ,
\end{equation}
where $E_{\gamma, \rm max}^{\rm VHE} \sim 0.1 \varepsilon'_{p, \rm max} \mathcal{D}_i/(1+z)$. Note that for a hard spectrum $L_\gamma \lesssim L_{\gamma, \rm VHE}$.
We estimate the VHE gamma-ray luminosity produced by photopion process using the relation $L_{\gamma, \rm VHE} \simeq 0.5 \, f_\pi \, L_p$.
Furthermore, in order to consider only potential TBL, we use the distribution of intrinsic spectral indexes to limit the contribution of BL Lacs to only those with spectral index from $\gamma_{\rm int,min} =1$ to $\gamma_{\rm int} \leq 2$; to do this, we use the spectral indexes distribution $dN/d\gamma_{\rm int}$
given by \cite{Zeng_2014MNRAS.441.1760Z}. 
Finally, the individual contribution of a source, considering all flavor neutrino flux, is given by the expression
\begin{equation}
    Q_\nu (E_\nu) 
    \simeq 
    \frac{3 f_\pi \mathcal{D}_i^2}
    {8 \left< \chi_{p\rightarrow \nu} \right>^2} 
    Q'_p(\varepsilon_p') \, ,
\end{equation}
where $\left< \chi_{p\rightarrow \nu} \right>\approx 0.05$ is mean the fraction of energy transfer from parent proton to the neutrino, the proton source if related with proton spectrum as $Q_p' \simeq 4\pi {R'}^2_b c N_p'$, and the observed neutrino energy is related with the proton energy as $\varepsilon_p' \simeq E_\nu (1+z)/(\left< \chi_{p\rightarrow \nu} \right>\mathcal{D}_i)$. It is important to mention that by inserting all these conditions into equation \ref{eq_diffuse_neutrino}, our result only depends on the value of $\varepsilon_{p,\rm max}'$, $\Gamma_i ( \mathcal{D}_i)$ and $L_{\gamma, \rm max}$.

We calculate the diffuse neutrino flux for different cases, considering $\varepsilon'_p= 100 \, \rm TeV$ and $\varepsilon'_p= 100 \, \rm PeV$ as boundary cases.  While the choice of $\Gamma_i (\mathcal{D})$ is taken from the results of modeling each TBL; Table \ref{tab_model_parameters} shows that $\Gamma_i$ takes values in the range of $\sim 1.5-3$.  Finally, we set $L_{\gamma, \rm max}^{\rm VHE} = 10^{45} \, \rm erg \, s^{-1}$, as the lower value.

\begin{figure}
\begin{minipage}[b]{\linewidth}
\centering
\includegraphics[width=\linewidth]{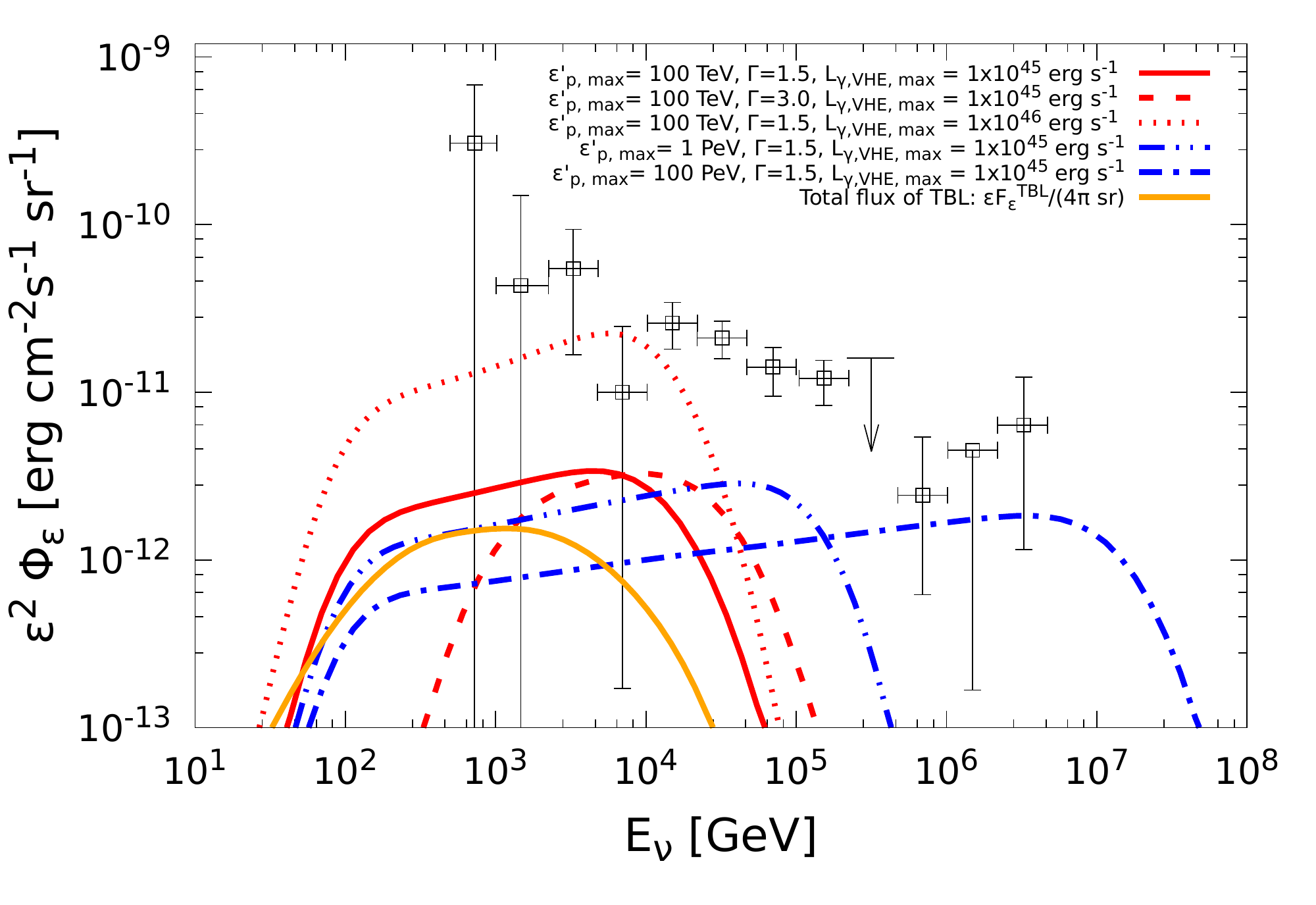} 
\end{minipage}
\caption{The diffuse neutrino flux observed by IceCube and neutrino flux estimated with our photo-hadronic model for hard-TeV BL Lac population considering different cases. We also show the total neutrino flux divided by $4\pi \rm \, sr$ calculated by summing up each individual source. The Icecube's data correspond to the neutrino spectrum of the shower like events \citep{IceCube2020arXiv200109520I}.}
\label{fig_Fneutrinos_diffuse}  
\end{figure}

\begin{figure*}
\begin{minipage}[b]{0.50\linewidth}
\centering
\includegraphics[width=\linewidth,height=.23\textheight]{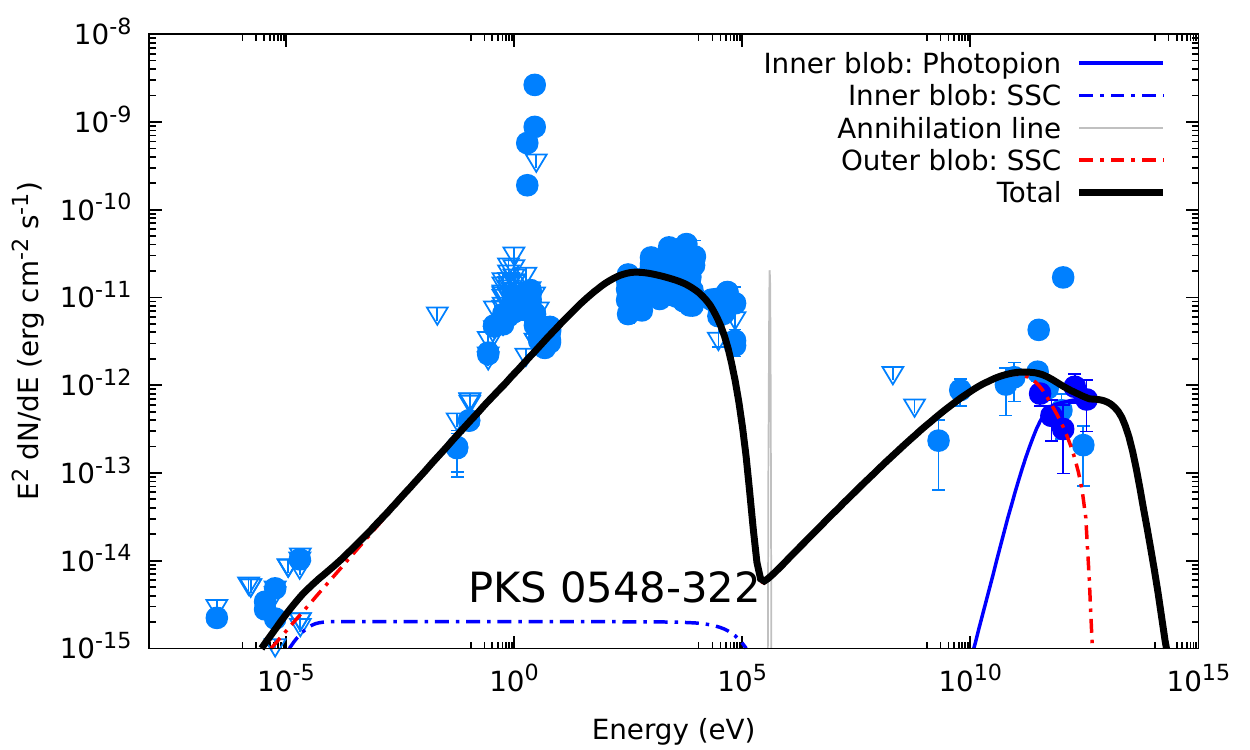}
\end{minipage}\hfill 
\begin{minipage}[b]{0.50\linewidth}
\centering
\includegraphics[width=\linewidth,height=.23\textheight]{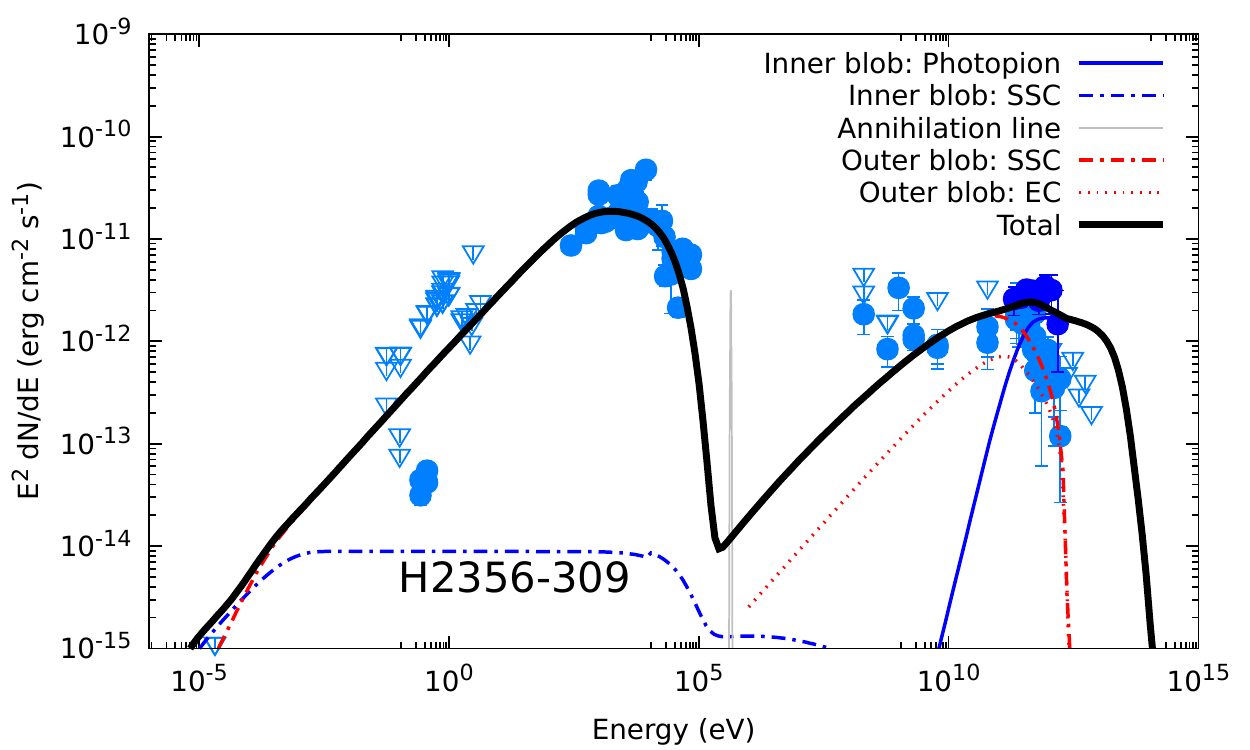} 
\end{minipage} 
\begin{minipage}[b]{0.50\linewidth}
\centering\includegraphics[width=\linewidth,height=.23\textheight]{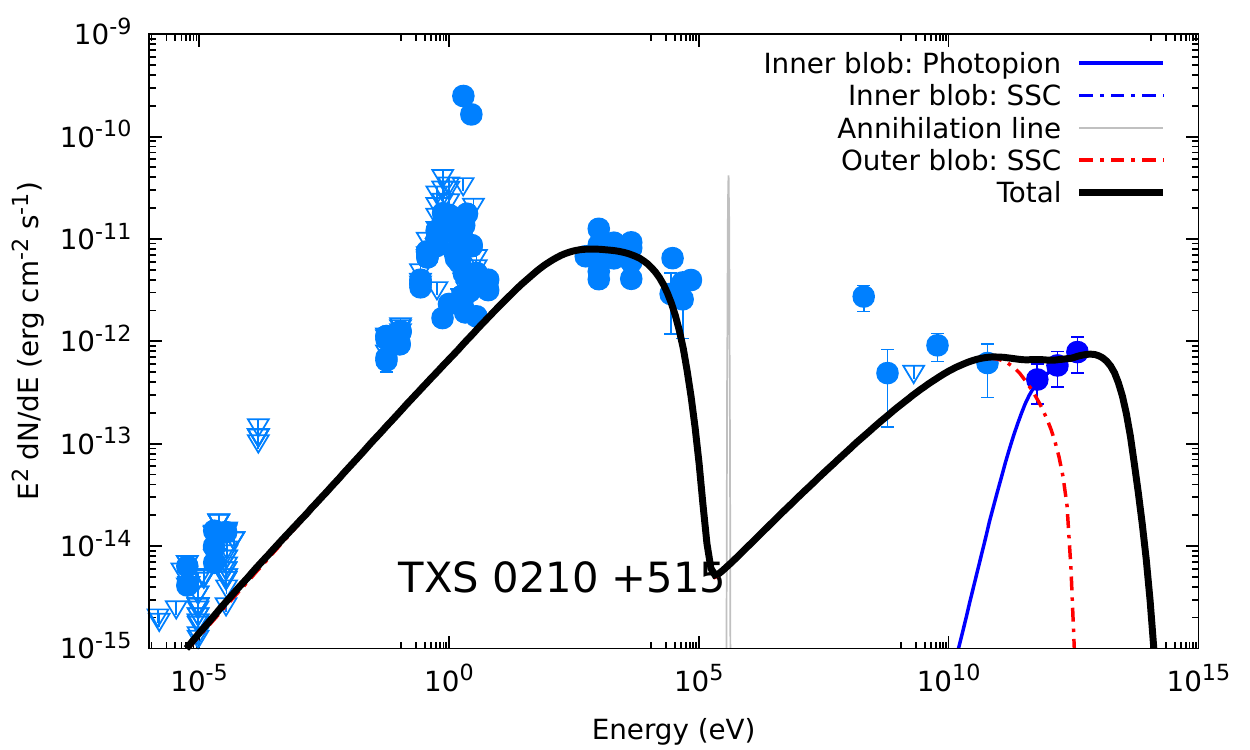}
\end{minipage}\hfill 
\begin{minipage}[b]{0.50\linewidth}
\centering\includegraphics[width=\linewidth,height=.23\textheight]{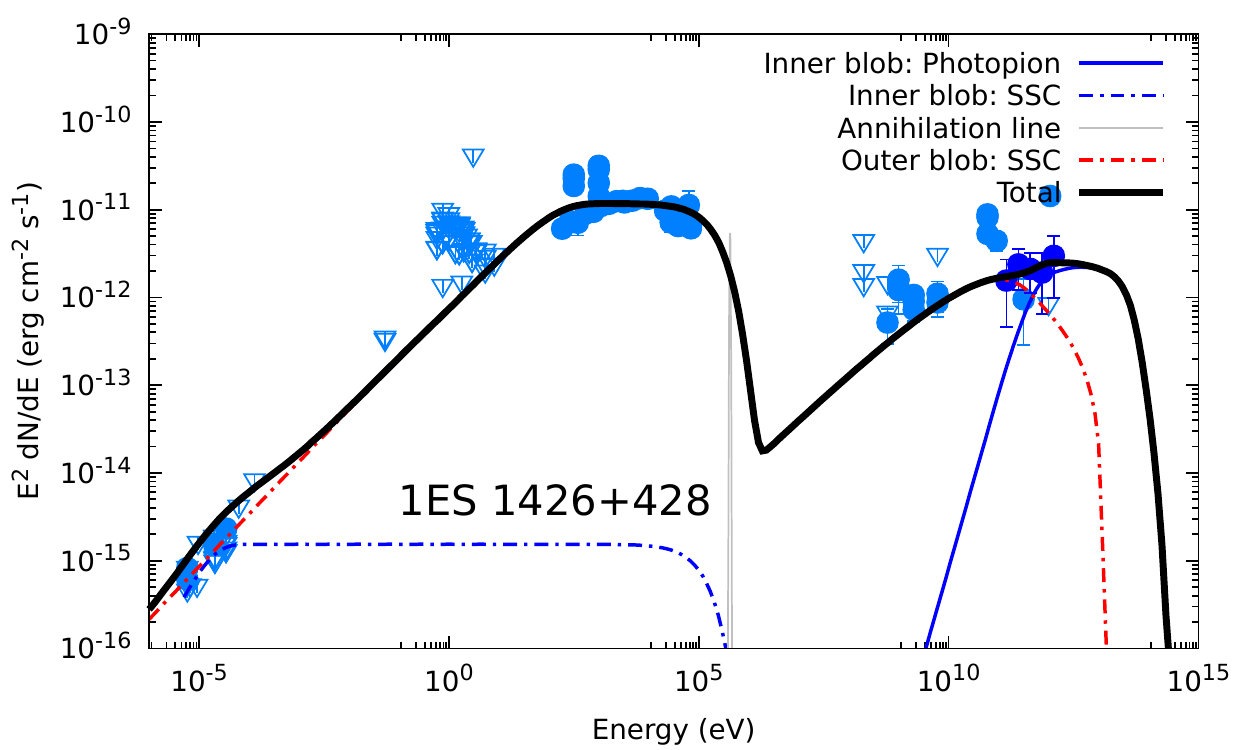}
\end{minipage} 
\begin{minipage}[b]{0.50\linewidth}
\centering\includegraphics[width=\linewidth,height=.23\textheight]{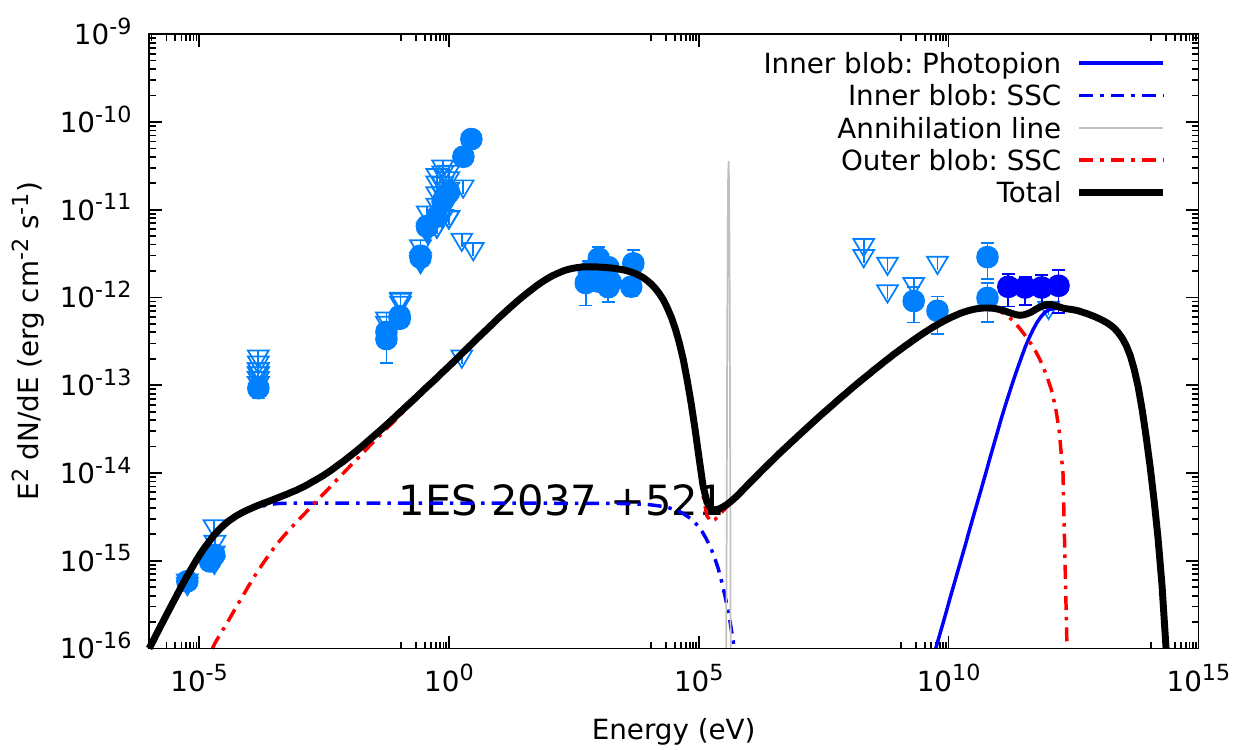}
\end{minipage}\hfill 
\begin{minipage}[b]{0.50\linewidth}
\centering\includegraphics[width=\linewidth,height=.23\textheight]{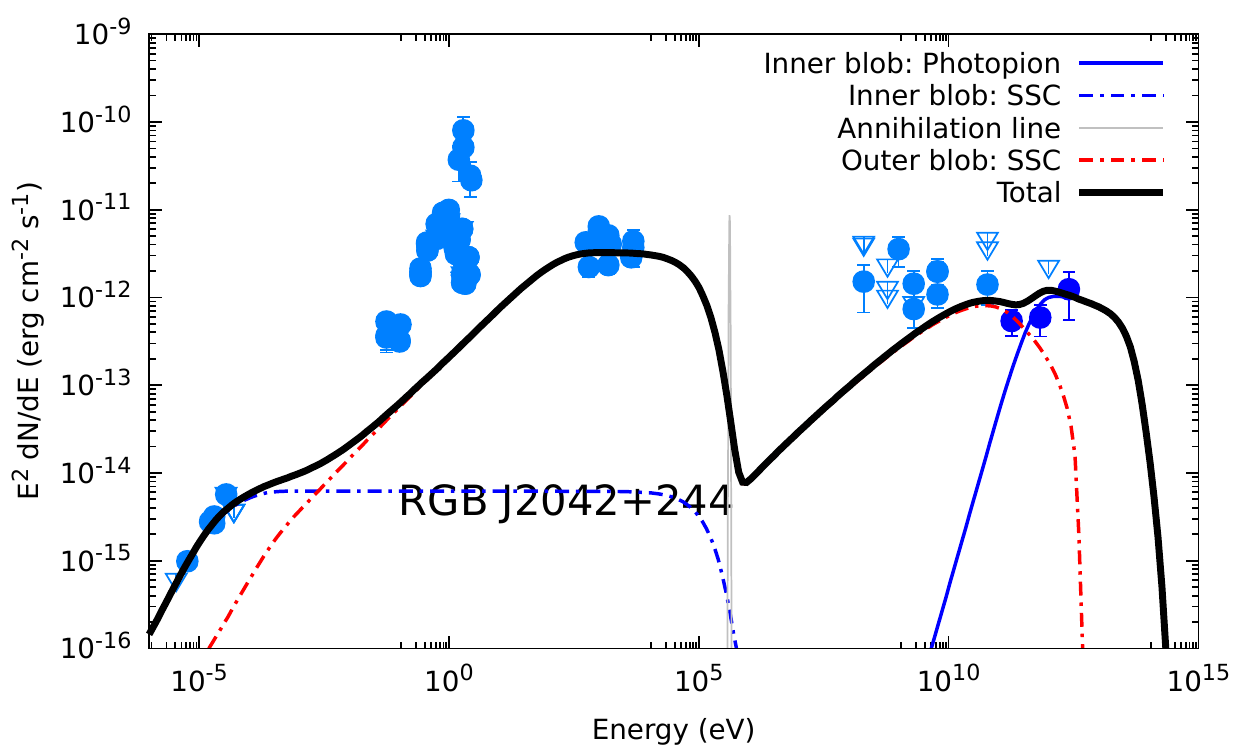}
\end{minipage}\hfill 
\begin{minipage}[b]{0.50\linewidth}
\centering\includegraphics[width=\linewidth,height=.23\textheight]{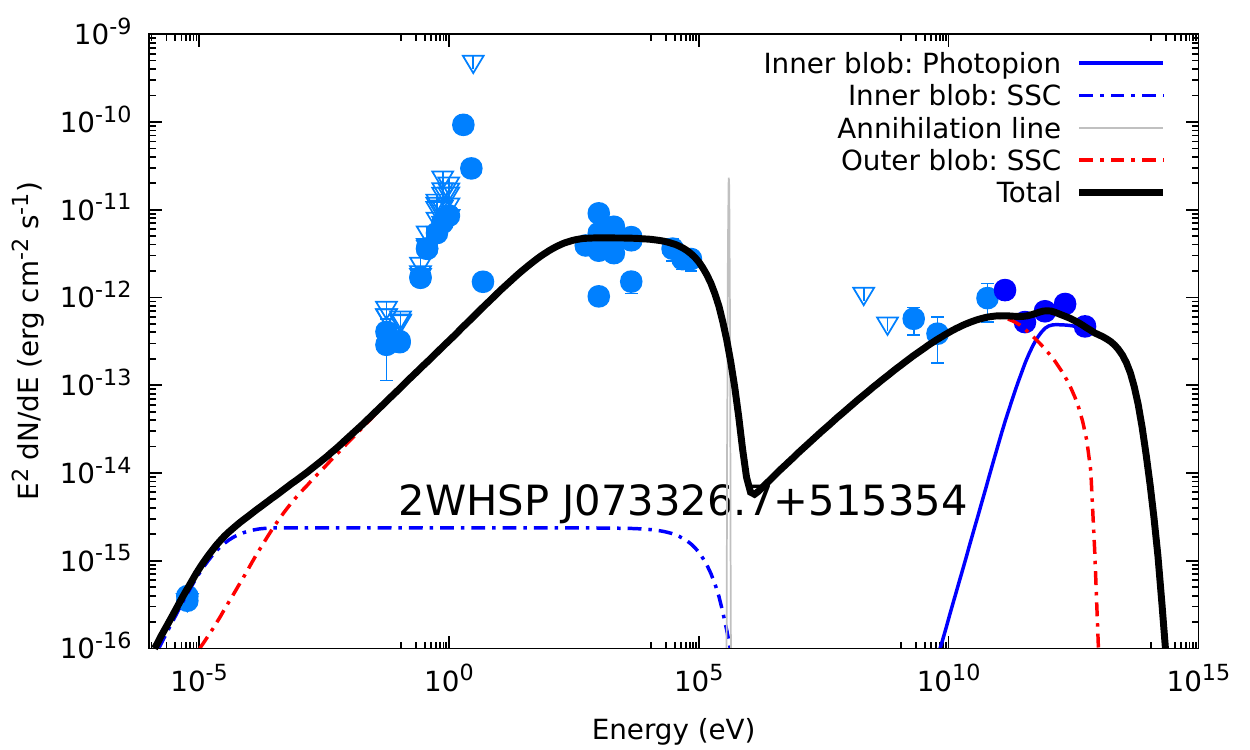}
\end{minipage}\hfill 
\begin{minipage}[b]{0.50\linewidth}
\centering\includegraphics[width=\linewidth,height=.23\textheight]{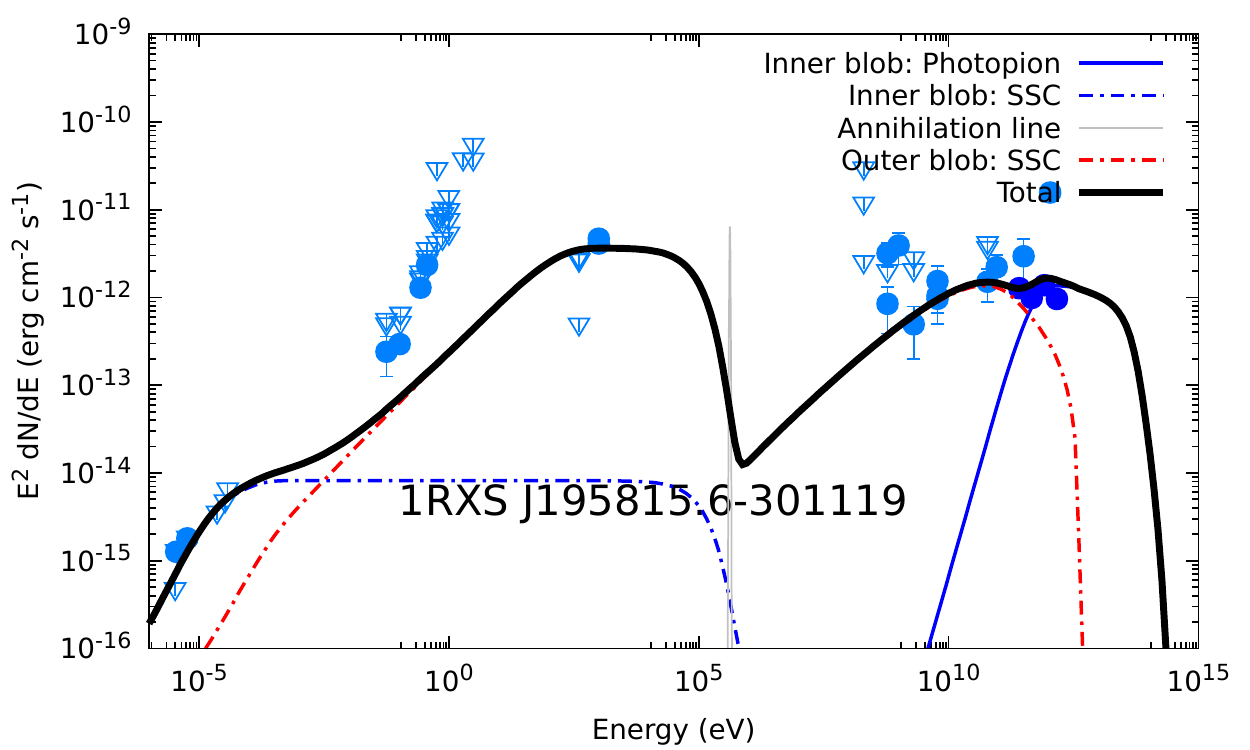}
\end{minipage}\hfill 
\caption{The broadband SED of eight  additional TBL calculated using our proposed model which was not included by  \citep{AguilarRuiz_2022MNRAS.512.1557A}. The archival data was taken from https://tools.ssdc.asi.it/SED/}
\label{fig_VHEspectrum}
\end{figure*}

Our result is plotted in Figure \ref{fig_Fneutrinos_diffuse}, compared with the diffuse neutrino flux observed by IceCube. It indicates a consistent flux level.  Moreover, we compare our result with the total flux estimated by the TBL catalog divided by $4\pi \, \rm sr$. This result indicates that the current TBL must dominate the neutrino flux detected for models with $L_{\gamma, \rm max}^{\rm VHE} = 10^{45} \, \rm erg \, s^{-1}$.
Note that the model with $\varepsilon'_{p, \rm max}= 100 \, \rm PeV$ is inconsistent with the total neutrino flux of the TBL population; therefore, we can rule out this case. On the other hand, the preferred models are those with $\Gamma_i=1.5$ and $L_{\gamma, \rm max}^{\rm VHE} \gtrsim 10^{45} \, \rm erg \, s^{-1}$. Furthermore, the model with $\varepsilon_p'= 1\, \rm PeV$ is the lowest flux allowed to be consistent with the TBL flux, but this implies that the current TBL population would dominate the total diffuse flux.
Note that our model with $\varepsilon'_{p, \rm max}= 100 \, \rm TeV$ does not predict HE-neutrinos either of the H.E.S.E. or muon-track neutrino catalogs, which have energies greater than 30 TeV. The total flux is of the order of $\sim 2 \times 10^{-12} \, \rm erg \, s^{-1} \, sr^{-1}$ between $\approx 0.1-10 \, \rm TeV$ and could represent only the one to ten percent of the observed diffuse flux at that energies (see figure \ref{fig_Fneutrinos_diffuse}).
It is important to mention that due to $f_\pi < 1$, surviving accelerated protons should reach the outer regions of the jet, and they can interact with the synchrotron emission in that regions or with photons of the BLR and the DT to produce neutrinos with higher energies \cite[e.g.,][]{Murase_2014PhRvD..90b3007M, Petropolou_2020ApJ...899..113P, Cerruti_2015_MNRAS.448..910C}.

If VHE gamma-rays of TBL have a hadronic origin, it could suggest that TBL cannot be the only source of neutrino emission at TeV energies. Another option could be the existing undetected population, implying a greater flux coming from TBL.
Here we consider only the TBL detected up to date. Moreover, it is essential to mention that \citep{Costamante2020MNRAS.491.2771C} enlisted 47 BL Lacs as candidates of hard-TeV BL Lacs that, if detected, could increase the neutrino flux at least by a factor of $\sim 3$ if they have a similar contribution as those considered in this work. These observations could test our result, comparing it with our prediction of diffuse neutrino flux. For instance, our model with the lowest values of $L_{\gamma, \rm max}^{\rm VHE} = 10^{45} \, \rm erg \, s^{-1}$ and $\varepsilon'_{p,\rm max}=100 \, \rm TeV$ is only a factor of $\sim 2$ greater than the flux obtained when TBL population is considered.


%
%
%
\section{Discussion and Conclusion}

We have modeled the SED of eight hard-TeV BL Lacs with our proposed leptohadronic scenario, in addition to the previous six modelled by \citep{AguilarRuiz_2022MNRAS.512.1557A}. Our model has two indispensable ingredients: (i) the presence of a pair-plasma emitting an annihilation line-spectrum centered at 511 keV and (ii) the existence of a blob that accelerates protons to energies at least to dozen or hundreds of TeV. Our model predicts a gamma-ray spectrum in the VHE band with an energy peak above $\sim \rm TeV$, consistent with the observations. We discussed the role of the secondary electrons population and pointed out that only photopion pairs are relevant at energies beyond $\gtrsim 1 \rm \, MeV$. It is worth noting that its signature peaks near the MeV band, where we lack data and cannot perform a correlation with gamma-rays. Our result of each TBL shows a good fit of our model with the archival data.

Another important result is that our model cannot account for any HESE or EHE neutrino event.   It is essential to mention that the detection of TBL is difficult due to the low flux level of these in the band of the Fermi-LAT. Thus a hidden TBL population could remain undetected if this population is there. To avoid overshooting, the observed neutrino flux and the number of undetected TBLs could not be greater than dozen-to-hundreds if we consider that they have gamma-ray contributions similar to those considered in this work. Furthermore, the diffuse neutrino flux was calculated using our model for TBL source. Our result agrees with the data reported by IceCube collaboration
by considering our scenarios presented in the Figure \ref{fig_Fneutrinos_diffuse}, only those with $\Gamma_i = 1.5$, $\varepsilon'_{p, \rm max}<1 \rm \, PeV$, and $L_{\gamma, \rm max}^{\rm VHE} > 10^{45} \, \rm erg \, s^{-1}$ agrees with the flux of the current TBL population.  In our model, the two latter parameters are not well constrain, for $\varepsilon'_{p,\rm max}$ the lower limit is $\sim 100 \, \rm TeV$, while for $L_{\gamma, \rm max}^{\rm VHE} < 10^{46} \, \rm erg \, s^{-1}$. If we consider the scenario with the lower values, our model predicts only a tiny fraction of one-to-ten percent of the flux observed in the energy range of $\sim$0.1-10 TeV. Our model is very sensitive to the parameter $L_{\gamma, \rm max}^{\rm VHE}$, but TBL observations could suggest the values could be close to $10^{45} \, \rm erg \, s^{-1}$ rather than $10^{46} \, \rm erg \, s^{-1}$. If this claim is right, from figure \ref{fig_Fneutrinos_diffuse}, we note that the current TBL should dominate the whole TBL population in the universe detected up to nowadays.

Future observatories such as Cherenkov Telescope Array could help discover the TBL population and help to constrain our model. Note that flare activities,  exhibited in some EHSP,  were not considered here, but they may be detected with neutrinos by futures instruments \cite[see,][for a recent discussion]{2022NatRP...4..697G}. In our model, MeV gamma-rays are suppressed, then could be tested with future observation in that band with future telescopes such as eASTROGAM or AMEGO.
If our proposed mechanism operates in that stage, it could be testable with neutrino observations with IceCube in the nearby blazars as Mrk 501 or Mrk 421. Finally, our model is sensitive to the maximum values of $\varepsilon_{p, \rm max}$ because they impact the peak of the neutrino spectrum. Therefore future neutrino observatories such as IceCube-gen2 and KM3Net are essential to the constraint that value.


\section{Acknowledgements}

We are grateful to Antonio Marinelli and Peter Veres for useful discussions.  This work is supported  by UNAM-DGAPA-PAPIIT  through  grant  IN106521.

%

\bibliography{apssamp}

\begin{table*} 
\caption{This table shows the complete TBL catalog used in this work. The first eight TBLs are listed in \citep{Costamante2020MNRAS.491.2771C} (see reference therein) and the new six were recently reported by \citep{MAGIC2019MNRAS.490.2284M,MAGIC2020ApJS..247...16A,2021arXiv210802232D}. Columns (2) and (3) are the coordinate given for right ascension and declination (data are taken from http://tevcat2.uchicago.edu/). Column (4) is the calculated luminosity distance. Column (5) is the redshift taken from (data are taken from http://tevcat2.uchicago.edu/). Column(6) is the lower limit of VHE peak energy. Column (7) is the lower limit of VHE luminosity at the energy peak, and Column (8) is the intrinsic VHE spectral slope.}
\label{tab_Obs_parameters}
\begin{tabular}{l l c c c c c c c c}
\hline
 Object name & R.A. & Dec &   $d_L$  & $z$ &  $E_{\gamma}^{pk}$ & $\varepsilon L_{\varepsilon_\gamma}^{pk}$ & $\gamma_{\rm VHE}^{\rm intr}$ & ref
 \\
  & (deg) & (deg) & $(\rm Gpc)$ &  & (TeV) & $(\rm erg \, s^{-1})$ & 
\\
  $[1]$ & [2] & [3] & [4] & [5] & [6] & [7] & [8] 
\\
\hline
\hline
\textbf{1ES 0229 +200 }  & 38.22 & 20.27 &  0.643 & 0.140 & $ > 12 $ & $ > 9.9 \times 10^{44}$ 
& $1.5 \pm 0.2$ & \citep{Aharonian_2007AA...475L...9A, Costamante2018,Costamante2020MNRAS.491.2771C}
\\
\textbf{RGB J0710 +591 } &  107.61 & 59.15 &   0.569 & 0.125 & $> 4 $ & $ > 2.3 \times 10^{45}$ 
& $1.8 \pm 0.2$ & \citep{Costamante2018,Costamante2020MNRAS.491.2771C}
\\
\textbf{1ES 0347-121}    &  57.35  & -11.97  &   0.899 & 0.188 & $> 3 $ & $ > 7.7 \times 10^{44}$ 
& $1.8 \pm 0.2$ & \citep{Costamante2018,Costamante2020MNRAS.491.2771C}
\\
\textbf{1ES 1101-232}    &  165.90 & -23.50  & 0.879 & 0.186 & $> 4 $ & $ > 9.2 \times 10^{44}$ 
& $1.7 \pm 0.2$ & \citep{Costamante2018,Costamante2020MNRAS.491.2771C}
\\
\textbf{1ES 0414 +009}   &  64.22 & 1.09    &  1.435 & 0.287 & $> 2 $ & $ > 7.3 \times 10^{44}$ 
& $1.9 \pm 0.3$ & \citep{Costamante2018,Costamante2020MNRAS.491.2771C}
\\
\textbf{1ES 1218 +304}   &  185.36 & 30.19 &  0.858 & 0.182 & $> 2 $ & $ > 1.8 \times 10^{45}$ 
& $1.9 \pm 0.1$ & \citep{Costamante2018,Costamante2020MNRAS.491.2771C}
\\

%
%
\textbf{PKS 0548-322 }$^a$ & 87.66 & -32.27 & 0.310  & 0.069 & $ > 2 $ & $ > 6.9 \times 10^{42}$ 
& $2.0 \pm 0.3$ & \citep{Ahanorian_2010AA...521A..69A,2008ICRC....3..913S}
\\
\textbf{H 2356-309 }$^a$   & 359.79 & -32.62 & 0.792 & 0.165 & $ > 2 $ & $ > 7.5 \times 10^{43}$ 
& $1.95 \pm 0.2$ & \citep{2012arXiv1208.0808C,2010A_A...516A..56H}
\\
%
%
\textbf{TXS 0210 +515}     & 33.57 & 51.75  & 0.217 & 0.049 & $> 4 $ & $ > 1.7 \times 10^{43}$ 
& $1.6 \pm 0.3$ & \citep{MAGIC2020ApJS..247...16A}
\\
\textbf{1ES 1426 +428}     & 217.14 & 42.70 & 0.605 & 0.129 & $>2  $ & $ > 1.3 \times 10^{43}$ 
& $1.8 \pm 0.5$ & \citep{MAGIC2020ApJS..247...16A}
\\
\textbf{1ES 2037 +521}$^{a}$  & 309.85 & 52.33 &  0.236   & 0.053 & $>2  $ & $ > 6.3 \times 10^{42}$ 
& $2.0 \pm 0.5$ & \citep{MAGIC2020ApJS..247...16A}
\\
\textbf{RGB J2042 +244}$^{b}$   & 310.53 & 24.45 & 0.480 & 0.104 & $>2  $ & $ > 5.5 \times 10^{42}$ 
& $1.8 \pm 0.6$ & \citep{MAGIC2020ApJS..247...16A}
\\
%
%
\textbf{2WHSP J073326.7+515354}$^a$ & 113.36 & 51.90 & 0.292 & 0.065 & $>2  $ & $ > 6 \times 10^{42}$ 
& $1.99 \pm 0.16$ & \citep{MAGIC2019MNRAS.490.2284M}
\\
\textbf{1RXS J195815.6-301119}$^a$ & 299.56 & -30.19 & 0.554 & 0.119 & $>2  $ & $ > 3.7 \times 10^{43}$ 
& $2.0 \pm 0.27$ & \citep{2021arXiv210802232D}
\\
\hline
\end{tabular}\label{tab_EHSP_list}
   \\
   \textbf{ Notes:} 
   \\
   $^a$ Due to the value on the spectral index, these objects could be considered as transitional soft-to-hard TeV BL Lac.
   \\
   $^b$ only a signal hint was reported \citep{MAGIC2020ApJS..247...16A}.
\end{table*}

\newpage

%
%
%


%
%
%

\clearpage


\begin{table*}


\caption{The best-fit, assumed and derived quantities found and used to describe the broadband SED of flares activities on Mrk 501 using the two-zone lepto-hadronic model proposed by \citep{AguilarRuiz_2022MNRAS.512.1557A}.}

\begin{adjustbox}{width=2.1\columnwidth}
\begin{tabular}{l c c c c c c c c c}
\hline\hline
  & \textbf{PKS0548-322} 
  & \textbf{H2356-309} 
  & \textbf{TXS0210+515}  
  & \textbf{1ES1476+428}  
  & \textbf{1ES2037+521}  
  & \textbf{RGBJ2042+244} 
  & \textbf{2WHSP} 
  & \textbf{1RXS }
\\
  & \textbf{ } 
  & \textbf{ } 
  & \textbf{ }  
  & \textbf{ }  
  & \textbf{ }  
  & \textbf{ } 
  & \textbf{J073326.7+515354} 
  & \textbf{J195815.6-301119}
\\
\hline
\hline \\
\textbf{INNER BLOB }\\
\\
{\bf Best-fit Quantities}\\
\textbf{$\Gamma_i $}      
& $1.52 \pm 0.15$ 
& $1.63 \pm 0.19$   
& $1.51 \pm 0.16$   
& $2.05 \pm 0.20$       %
& $2.11 \pm 0.25$     %
& $2.03 \pm 0.19$       %
& $2.11 \pm 0.24$  		
& $2.23 \pm 0.25$ 		%
\\
\textbf{$R_b$}[$10^{14} \, \rm cm$]  
& $2.03 \pm 0.21$    
& $5.05 \pm 0.51$    
& $0.82 \pm 0.1$  
& $2.13\pm 0.18$ %
& $1.35\pm 0.12$ 
& $1.92\pm 0.16$ 
& $1.23\pm 0.10$ 
& $2.14\pm 0.20$ 
\\
\textbf{$B \; \rm [10^{2}\,G]$ }   
& $0.31\pm 0.04$  
& $0.11\pm 0.01$  
& $0.30\pm 0.03$  
& $0.21\pm 0.02$  
& $0.22\pm 0.02$  
& $0.20\pm 0.02$  
& $0.21\pm 0.02$  
& $0.20\pm 0.02$  
\\
$K_p[\rm 10^{-4}\,eV^{-1} \, cm^{-3}]$
    & $0.87 \pm 0.01$  
    & $1.51  \pm 0.12$   
    & $0.62  \pm 0.08$   
    & $1.54  \pm 0.15$  
    & $3.65  \pm 0.44$  
    & $2.51  \pm 0.31$   
    & $3.51  \pm 0.30$  %
    & $4.41 \pm 0.57$ \\ 
\\
{\bf Assumed Quantities}\\
$\varepsilon_{p,\rm max} [\rm PeV]$ \cite[e.g.,][]{AguilarRuiz_2022MNRAS.512.1557A}
    & $0.1$  
    & $0.1$  
    & $0.1$   
    & $0.1$   
    & $0.1$    
    & $0.1$    
    & $0.1$     
    & $0.1$     
\\
\textbf{$\gamma_{e,\rm b}$}\cite[e.g.,][]{Cerruti_2015_MNRAS.448..910C}   
    & $1$      
    & $2$      
    & $5$      
    & $1.3$      %
    & $2.5$      
    & $1.3$		
    & $2.5$		
    & $1.3$		
\\
\textbf{$\alpha_{\rm e,1}=\alpha_p$} \cite[e.g.,][]{Costamante2018,Costamante2020MNRAS.491.2771C}
& 1.9  
& 1.95 
& 1.6  
& 1.8  %
& 2  %
& 2  
& 2  
& 2  
\\
\textbf{$\alpha_{\rm e,2}$} \cite[e.g.,][]{Cerruti_2015_MNRAS.448..910C}
& 3
& 3
& 3
& 3
& 3 
& 3 
& 3 
& 3 
\\
\\
{\bf Derived Quantities}\\
$L_p\rm \; [10^{44}\,erg \, s^{-1}]$
    & $0.65 \pm 0.02$   
    & $3.70 \pm 0.71$   
    & $0.43 \pm 0.03$   
    & $5.10 \pm 0.16$  
    & $0.84 \pm 0.16$  
    & $2.40 \pm 0.27$   
    & $0.85 \pm 0.18$  
    & $4.20 \pm 0.93$   
\\
\textbf {$n_p=n_e\rm \; [10^{5}\,cm^{-3}]$}
    & $0.89 \pm 0.19$    
    & $1.50 \pm 0.29$    
    & $0.81 \pm 0.09$ 
    & $1.70 \pm 0.38$ %
    & $3.30 \pm 0.68$ 
    & $2.30 \pm 0.50$  
    & $3.30 \pm 0.64$  
    & $4.70 \pm 0.88$  
\\
\textbf{$L_e \rm \; [10^{39}\,erg \, s^{-1}]$}
    & $2.70 \pm 0.63$  
    & $32.21 \pm 6.10$  
    & $0.61 \pm 0.11$  
    & $18.12 \pm 3.80$ %
    & $9.60 \pm 1.80$ %
    & $23.10 \pm 4.30$  
    & $9.60 \pm 2.20$  
    & $41.1 \pm 8.10$  
\\
\textbf{$L_B\rm \; [10^{43}\,erg \, s^{-1}]$}
    & $6.32 \pm 1.39 $  
    & $4.53 \pm 0.85 $  
    & $9.33 \pm 2.31 $  
    & $9.71 \pm 2.18 $ %
    & $2.50 \pm 0.51 $ %
    & $9.16 \pm 1.81 $  
    & $2.91 \pm 0.46 $ 
    & $9.43 \pm 2.15 $ 
\\
\textbf{$U_B/(U_p+U_e)$}
& $0.96 \pm 0.17$   
& $0.12 \pm 0.03$   
& $0.20 \pm 0.04$ 
& $0.30 \pm 0.06$ %
& $0.29 \pm 0.06$ %
& $0.40 \pm 0.08$   
& $0.28 \pm 0.06$   
& $0.23 \pm 0.05$   
\\
\\
\hline \\
%
\textbf{OUTER BLOB } \\
\\
{\bf Best-fit Quantities}\\
\textbf{$\Gamma_o$}      
& $5.11 \pm 1.16$  
& $5.23 \pm 1.13$ 
& $5.22 \pm 1.13$   
& $5.01 \pm 1.10$   
& $5.01 \pm 1.11$   %
& $5.61 \pm 1.21$ 
& $10.91 \pm 2.31$ 
& $5.70 \pm 1.24$ 
\\
\textbf{$B \rm \; [G]$}     
& $0.19\pm 0.03$ 
& $0.35\pm 0.07$ 
& $0.34\pm 0.07$ 
& $0.13\pm 0.03$ %
& $0.45\pm 0.08$ %
& $0.53\pm 0.12$  
& $0.46\pm 0.11$  
& $0.63\pm 0.14$  
\\
\textbf{$R_b$ [$10^{16} \, \rm cm$]} 
& $2.16\pm 0.52$ 
& $1.21\pm 0.28$ 
& $0.61\pm 0.11$ 
& $2.22\pm 0.51$ 
& $0.15\pm 0.02$ %
& $0.23\pm 0.03$ 
& $0.11\pm 0.02$ 
& $0.22\pm 0.02$ 
\\
\textbf{$\alpha_{\rm e,1}$} 
& $1.84\pm 0.41$  
& $1.93\pm 0.39$  
& $1.92\pm 0.33$  
& $1.82\pm 0.35$  %
& $1.86\pm 0.34$  %
& $1.84\pm 0.39$  
& $1.79\pm 0.33$  %
& $1.82\pm 0.34$  
\\
\textbf{$\alpha_{\rm e,2}$} 
& $3.31 \pm 0.61$ 
& $3.22 \pm 0.62$
& $3.11 \pm 0.67$
& $3.01 \pm 0.59$%
& $3.08 \pm 0.57$%
& $3.19 \pm 0.52$
& $3.01\pm 0.54$%
& $3.21 \pm 0.56$

\\
\\
{\bf Assumed Quantities}\\
\textbf{$\gamma_{\rm min}$} \cite[e.g.,][]{2017APh....89...14F}  
& $1$ 
& $50$ 
& $1$ 
& $1$ %
& $50$ %
& $50$ 
& $50$ %
& $50$ 
\\
\\
{\bf Derived Quantities}\\
\textbf{$\gamma_{\rm b}\,[10^{4}]$}   
    & $12.22 \pm 2.62$ 
    & $13.11 \pm 2.73$ %
    & $7.13 \pm 1.41$ 
    & $11.14 \pm 2.40$ %
    & $5.01 \pm 1.04$ %
    & $4.91 \pm 1.02$ 
    & $4.04 \pm 0.85$ %
    & $4.80 \pm 0.95$ 
\\
\textbf{$\gamma_{\rm max}\,[10^{6}]$} 
    & $1.51 \pm 0.31$ 
    & $0.63 \pm 0.11$ %
    & $0.69 \pm 0.10$ 
    & $3.22 \pm 0.73$ %
    & $53.14 \pm 12.32$ %
    & $1.61 \pm 0.35$ 
    & $3.51 \pm 0.73$ %
    & $3.12 \pm 0.71$ 
\\
\textbf{$L_B$} $\rm \; [10^{42}\,erg \, s^{-1}]$
    & $23.54  \pm 5.12 $  
    & $65.22 \pm 11.83$  
    & $5.51 \pm 1.26$ 
    & $23.11  \pm 5.63$  %
    & $0.48  \pm 0.12$  %
    & $3.31  \pm 0.67$  
    & $5.43  \pm 1.09$  
    & $3.71  \pm 0.73$  
\\
\textbf{$L_e$} $\rm \; [10^{43}\,erg \, s^{-1}]$
    & $5.43\pm 1.13$ 
    & $7.61\pm 1.76 $ 
    & $2.31\pm 0.49 $ %
    & $18.13\pm 4.11$ %
    & $1.32\pm 0.23$ %
    & $2.51\pm 1.53$ 
    & $7.72\pm 1.73$ 
    & $3.31\pm 0.63$ 
\\
\textbf{$U_B/U_e$}  
& $0.43\pm 0.08$ 
& $0.89\pm 0.20$ 
& $0.19\pm 0.03$ %
& $0.22\pm 0.04$ %
& $0.03\pm 0.005$ %
& $0.14\pm 0.03$ 
& $0.08\pm 0.01$ 
& $0.09\pm 0.02$ 
\\
\hline
\hline
\end{tabular} \label{tab_model_parameters}
\end{adjustbox}
\label{tab_model_parameters}
\end{table*}

\end{document}